\documentclass[11pt,a4paper,english,twoside]{article}

\usepackage{a4wide}
\usepackage{amssymb, amsmath, amsthm}
\usepackage{graphicx}
\usepackage{subcaption}
\usepackage[all]{xy}
\usepackage[pdftex,hyperref,svgnames]{xcolor}
\usepackage[pdftex,colorlinks=true,
pdfstartview=FitV,
pdfnewwindow=true,
linktoc = page,
linkcolor= blue,
citecolor= red,
urlcolor= blue,
hyperindex=true,
hyperfigures=false]{hyperref}
\hypersetup{linktocpage}
\usepackage{dsfont}
\usepackage{empheq}
\usepackage{cite}
\usepackage{float}
\usepackage{cancel}
\usepackage{relsize}
\usepackage{soul}
\usepackage{enumerate}
\usepackage{enumitem}
\usepackage{hhline}
\usepackage{multirow}
\usepackage{xspace}
\usepackage{bbm}
\usepackage{pdfpages}
\usepackage{setspace}
\usepackage{comment}

\newcommand{\nn}{\nonumber}

\newcommand{\be}{\begin{equation}}
\newcommand{\ee}{\end{equation}}
\newcommand{\ba}{\begin{eqnarray}}
\newcommand{\ea}{\end{eqnarray}}

\def\beq{\begin{equation}}
\def\eeq{\end{equation}}
\def\bea{\begin{eqnarray}}
\def\eea{\end{eqnarray}}

\def\del {\partial}
\def\d {{\rm d}}

\begin{document}
\numberwithin{equation}{section}

\begin{titlepage}
\begin{center}

\phantom{DRAFT}

\vspace{1.8 cm}

{\LARGE \bf{Single versus multifield scalar potentials \\ \vspace{0.3cm} from string theory}}\\

\vspace{2.0 cm} {\Large David Andriot$^{1}$, Muthusamy Rajaguru$^{2}$, George Tringas$^{2}$}\\

\vspace{0.9 cm} {\small\slshape $^1$ Laboratoire d'Annecy-le-Vieux de Physique Th\'eorique (LAPTh),\\
CNRS, Universit\'e Savoie Mont Blanc (USMB), UMR 5108,\\
9 Chemin de Bellevue, 74940 Annecy, France}\\

\vspace{0.4 cm} {\small\slshape $^2$ Department of Physics, Lehigh University,\\
16 Memorial Drive East, Bethlehem, PA 18018, USA}\\

\vspace{0.9cm} {\upshape\ttfamily andriot@lapth.cnrs.fr; mur220@lehigh.edu; georgios.tringas@lehigh.edu}\\

\vspace{2.2cm}
{\bf Abstract}
\vspace{0.1cm}
\end{center}
\begin{quotation}
In this work, we investigate the properties of string effective theories with scalar field(s) and a scalar potential. We first claim that in most examples known, such theories are {\sl multifield}, with at least 2 non-compact field directions; the few counter-examples appear to be very specific and isolated. Such a systematic multifield situation has important implications for cosmology. Characterising properties of the scalar potential $V$ is also more delicate in a multifield setting. We provide several examples of string effective theories with $V>0$, where the latter admits an asymptotically flat direction along an off-shell field trajectory: in other words, there exists a limit $\hat{\varphi} \rightarrow \infty$ for which $\frac{|\partial_{\hat{\varphi}} V|}{V} \rightarrow 0 $. It is thus meaningless to look for a lower bound to this single field quantity in a multifield setting; the complete gradient $\nabla V$ is then better suited. Restricting to on-shell trajectories, this question remains open, especially when following the steepest descent or more generally a gradient flow evolution. Interestingly, single field statements in multifield theories seem less problematic for $V<0$.
\end{quotation}

\end{titlepage}

\tableofcontents

\section{Introduction, summary and outlook}\label{sec:intro}

\subsection{Introduction}

The swampland program \cite{Vafa:2005ui, Palti:2019pca, vanBeest:2021lhn, Agmon:2022thq} aims at characterising the outcomes of a quantum gravity theory, such as its solutions or effective theories, and distinguishing them from models that cannot have a quantum gravity U.V.~completion. Of particular interest have been $d$-dimensional theories of scalar fields $\varphi^i$ minimally coupled to gravity, with $d\geq3$, since those commonly appear as effective theories of $10$-dimensional (10d) string theory compactifications and generalisations thereof (M- and F-theory constructions, Landau-Ginzburg models, non-geometric constructions, etc.). Such $d$-dimensional models have an action of the form
\beq
{\cal S} = \int \d^d x \sqrt{|g_d|} \left( \frac{M_p^{d-2}}{2} {\cal R}_d - \frac{1}{2} g_{ij}\, \del_{\mu} \varphi^i \del^{\mu} \varphi^j - V(\varphi^k) \right) \ , \label{action}
\eeq
where $g_{ij}(\varphi^k)$ stands for the field space metric, assumed definite positive, and $V(\varphi^k)$ is the scalar potential. Diffeomorphisms of field space may allow to reach a canonical basis of fields where $g_{ij} = \delta_{ij}$, in which case the canonically normalized fields are denoted as $\hat{\varphi}^i$. While such models are easily obtained from string theory, the characterisation question is about the details of $g_{ij}$ and $V$: in particular, {\sl can one state a generic property of a scalar potential $V$ obtained from string theory?}

Knowing features that are universal in string effective theories may have an important impact for phenomenology. Indeed, the models \eqref{action} in $d=4$ are relevant to cosmology, either as an inflation model, or to describe dark energy today via quintessence, or as a cosmological constant. The latter case would correspond to constant scalar fields and a constant potential value, which gets realised in a solution at an extremum of the scalar potential $V$. In that case, the spacetime is de Sitter. Is it possible to obtain such a de Sitter solution, or an observationally valid quintessence or inflation model, in the case where \eqref{action} is a string effective theory, or are there some obstructions? Having a characterisation, or knowing general features of $V$ should help answering this question. A lot of activity has been devoted in the last decade to such investigations, and characterising string effective theories of the form \eqref{action} and their potentials is again the topic of this work.\\

The de Sitter swampland conjecture is a proposal for a quantitative characterisation of string effective theories of the form \eqref{action} with $V>0$. Its initial version \cite{Obied:2018sgi} proposed that the following property should hold
\beq\label{covd}
\frac{\nabla V}{V} \geq c \ , \quad c \sim {\cal O}(1) \ ,
\qquad
{\rm with}
\quad
\nabla V \equiv \sqrt{g^{ij}\, \del_i V\, \del_j V} \ ,
\eeq
where $\del_i \equiv \del_{\varphi^i}$, and we set the reduced Planck mass $M_p$ to 1, considering from now on Planckian units. Although highly debated and soon refined, the condition \eqref{covd} carries several important features that will remain in the following. First, this proposed characterisation of scalar potentials with string theory origin is formulated {\sl off-shell}: \eqref{covd} gives a property of the potential $V$ (with $g_{ij}$) as a function, without referring to any physical solution of the theory. In other words, this condition is supposed to be valid anywhere in field space, instead of e.g.~restricting to physical trajectories; the latter would correspond to an {\sl on-shell} characterisation. Second, the property is formulated in a field space diffeomorphism-covariant manner, thanks to the object $\nabla V$. In particular, the condition invokes all fields and all corresponding derivatives. This is to be contrasted with single field conditions that we will discuss below. Finally, $\nabla V$ can be viewed as the norm of the gradient vector $\overrightarrow{\nabla}V$ of component $g^{ij} \del_i V$: $\nabla V = || \overrightarrow{\nabla}V ||  $. The condition \eqref{covd} can thus be viewed as characterising the overall slope of the potential at a given point in field space, requiring in addition a non-zero lower bound to it. Famously, this forbids vanishing slopes, $\nabla V =0$, which would correspond to a critical point or extremum of the potential, i.e.~a de Sitter solution.

Various refinements of the initial condition \eqref{covd} have been considered \cite{Andriot:2018wzk, Garg:2018reu, Ooguri:2018wrx, Andriot:2018mav, Rudelius:2019cfh, Bedroya:2019snp, Rudelius:2021oaz}, eventually leading to proposing such a claim only in the asymptotics of field space. By this, one means to consider (any) field limit $\hat{\varphi} \rightarrow \infty$ and characterising the potential there in a similar way as \eqref{covd}. Among these developments, it is important to note that many reasonings, proposals or checks have been based on single field examples, sometimes restricting to an exponential potential (common in the asymptotics of string theory examples). This typically led to consider the {\sl single field slope ratio} $\frac{|\del_{\hat{\varphi}} V|}{V}$ for a canonically normalized field. For example, the (initial version of the) Trans-Planckian Censorship Conjecture (TCC) \cite{Bedroya:2019snp} led to the following single field characterisation in $d=4$
\beq
\hat{\varphi} \rightarrow \infty: \qquad \text{TCC bound}:
\ \ \frac{|\del_{\hat{\varphi}} V|}{V} \geq \sqrt{\frac{2}{3}} \ ,\label{TCC}
\eeq
where the TCC value for $c$ becomes $c= 2/\sqrt{(d-1)(d-2)}$ for arbitrary $d\geq 4$. Turning to a multifield setting, this condition got straightforwardly generalised, considering a trajectory in field space parameterised by $\hat{s}$ together with the limit $\hat{s} \rightarrow \infty$ \cite{Bedroya:2019snp}. In addition, since in that case, one has $\nabla V \geq |\del_{\hat{s}} V|$, the condition \eqref{TCC} implies one formulated with $\nabla V$ (see also \cite[Sec. 4]{Andriot:2022brg} for discussions on the generalisation to multifield). The single field focus seems therefore harmless at this stage.

When restricting to a single field exponential potential, further results could be derived in \cite{Bedroya:2019snp}. It is well-known that such a setting can be treated as a dynamical system and solutions (or physical trajectories in field space) admit fixed points (see \cite{Andriot:2024jsh} for a recent, comprehensive, account). When considering only the scalar field and its potential as the physical content, the system admits a stable (i.e.~attractive) fixed point for $\frac{|\del_{\hat{\varphi}} V|}{V} < \sqrt{6}$, named $P_{\phi}$ in \cite{Andriot:2024jsh}. Applying the TCC at this asymptotic solution, one obtains the condition $\frac{|\del_{\hat{\varphi}} V|}{V} \geq \sqrt{2}$. More generally, considering generic positive scalar potentials in a multifield setting, and using different arguments (invariance under dimensional reduction), the same $c$ value was obtained in the so-called Strong de Sitter conjecture (SdSC) \cite{Rudelius:2021oaz, Rudelius:2021azq}. The resulting potential characterisation is given as follows in asymptotics of field space
\beq
\hat{\varphi} \rightarrow \infty: \qquad \text{SdSC bound}:
\ \ \frac{\nabla V}{V} \geq \sqrt{2} \ , \label{SdSC}
\eeq
where the $c$ value generalises to $c=2/\sqrt{d-2}$ in arbitrary dimension. It was noted that the two claims \eqref{TCC} and \eqref{SdSC} are not necessarily incompatible, since the former could be completed to the latter by considering (non-flat) extra scalar fields.

It is important to note that while the derivation or arguments leading to these claims may have used equations of motion, or some further aspects of physical trajectories in field space, they were often interpreted as {\sl off-shell} characterisations, i.e.~valid in any asymptotic. It was also the case for the numerous checks that have been carried out on those. To start with, there is up-to-date no known counter-example to the SdSC and the $d=4$ value $\sqrt{2}$. For instance, the presence of a rolling $d$-dimensional dilaton, as in perturbative (geometric) limits with only positive terms in the potential, is known to saturate already by itself the SdSC bound (see e.g.~\cite{Shiu:2023fhb}). Turning to the TCC single field condition, many no-go theorems against de Sitter were reformulated as in \eqref{TCC}, allowing to extract a $c$ value: interestingly, the TCC value was then often matched, and if not, a higher one did, in dimensions $d \geq 4$ \cite{Andriot:2020lea, Andriot:2022xjh}, offering striking checks of this proposal. Note that these checks involved one field that was a combination of the dilaton and volumes in the internal, $(10-d)$-dimensional, compactification manifold.

However, the work \cite{Calderon-Infante:2022nxb} found a counter-example to the TCC single field condition \eqref{TCC}. Focusing on Calabi-Yau compactifications with fluxes, in F-theory or in type IIB supergravity, and restricting to the complex structure moduli space (with axio-dilaton), one field limit was identified for which $c= \sqrt{\frac{2}{7}} < \sqrt{\frac{2}{3}}$. The discrepancy with the above checks can be explained by the fact that internal volumes require the K\"ahler moduli, ignored in \cite{Calderon-Infante:2022nxb}; in other words, the complex structure sector alone had not been probed by the previous checks. Note that when adding the K\"ahler moduli, to build the complete $\nabla V$, the SdSC is still expected to hold. Indeed, the de Sitter no-go theorem worked-out for ${\cal R}_6 \geq 0$ (see e.g.~\cite[Sec. 4.2.4]{Andriot:2022xjh}), applicable here for a Calabi-Yau compactification with fluxes, and that involves the volume field, gives a $c$ value higher than $\sqrt{2}$.\\

The characterisation of (positive) string theory scalar potentials described above raises several questions. While the SdSC \eqref{SdSC} {\sl off-shell} and multifield covariant condition is so far verified, an {\sl off-shell} single field condition as expressed here in \eqref{TCC} is subject to doubt. We then ask the following question, while restricting to $V>0$
\begin{equation}
\begin{split}
& \text{{\sl For a string theory example, in the asymptotics of a multifield space,}} \label{question1}\\
& \text{{\sl what is the (lowest) bound $c$ in a single field condition}}\ \frac{|\del_{\hat{\varphi}} V|}{V} \geq c\ \text{{\sl ?}}
\end{split}
\end{equation}
where we recall that $\hat{\varphi}$ is canonically normalized. Note that this question is asked {\sl off-shell}. The emphasis on this point, and the fact that some of the derivations or checks mentioned were done {\sl on-shell}, raises in addition the following question:
\bea
& \text{{\sl For a string theory example, in a multifield space,}}\nn\\
& \text{{\sl if one restricts to physical trajectories,}} \label{question2}\\
& \text{{\sl what (on-shell) characterisation of the scalar potential can be made?}}\ \nn
\eea
This question deserves several comments that we will come back to in Section \ref{sec:onshell}, but we can already note that physical trajectories are a restriction of the whole field space. Therefore, the SdSC \eqref{SdSC} should still hold when considered {\sl on-shell}.\\

Distinctions made above between {\sl off-shell} and {\sl on-shell} conditions, and single versus multifield models, are crucial and related: indeed, in a single field example, there is no {\sl off-shell} notion, because there is only one trajectory that can be explored in the potential, which therefore matches the physical one; the only exception might be in the case of a vacuum, that may forbid a solution to explore the asymptotics. Departure from purely single field examples and reasonings is therefore where ambiguities start, hence the motivation to answer the above questions.\footnote{\label{foot:asymptotic}One ambiguity in a multifield situation is the notion of asymptotics. Considering the limit in, say, $\hat{\varphi}^1 \rightarrow \infty$ of a multivalued function $V(\hat{\varphi}^i)$ is a priori possible. But what is often implicitly considered is rather to follow a (single field) trajectory along $\hat{\varphi}^1$. In a multifield case, a trajectory (a 1-dimensional subspace) is not defined only by its direction $\hat{\varphi}^1$, but also by the values taken by the transverse coordinates, meaning by indicating the point locus in the transverse hyperplane. The latter is not often specified; we will do so in our investigations. However, claims made ``for any asymptotic'' do not need those specifications.}

\subsection{Results summary}

A good starting point is to discuss, as done in Section \ref{sec:singlefield}, the possibility of getting a single field model from string theory. We argue that a truly single field effective theory in $d\leq 9$ is very uncommon: known examples realising this are very few, specific and seemingly isolated. We summarize this situation as follows
\bea
& \text{{\bf Field Space Statement:}} \label{conj}\\[2mm]
& \text{{\sl In a $d$-dimensional effective theory of string theory, $3 \leq d \leq 9$, in most examples known,}} \nn\\
& \text{{\sl the scalar field space has real dimension at least equal to 2, i.e.~is multifield.}} \nn\\
& \text{{\sl In addition, at least 2 of its dimensions are non-compact.}}\nn
\eea
Let us emphasize that the 2 ``non-compact'' scalars, just mentioned to be almost systematically present, do not include ``pseudo-scalars'', e.g.~axions, since the latter have a ``compact'' field space (in absence of background flux). Also, we implicitly refer here to neutral scalar fields, i.e.~uncharged with respect to possible gauge groups. This statement applies to moduli spaces, that have been studied a lot, and more generally to field spaces of e.g.~effective theories of the form \eqref{action}.

As discussed in Section \ref{sec:singlefield}, the intuition behind this statement is that for $3 \leq d \leq 9$, one gets through standard geometric dimensional reductions both the dilaton and the internal volume as scalar fields, and in less geometric derivations, one typically gets the dilaton and some complex structure modulus: this gives indeed 2 non-compact scalars. It is therefore in very specific non-geometric constructions \cite{Dabholkar:1998kv, Baykara:2023plc, Baykara:2024vss, Baykara:2024tjr}, where this intuition breaks down, that this statement does not fully hold. As will be detailed in Section \ref{sec:singlefield}, those few and tailored examples include various kinds of asymmetric orbifolds, which have in their $d$-dimensional massless spectrum only one non-compact neutral scalar. We will also discuss there the case of M-theory, which should be considered with care in that respect.

The above statement extends a ``lower bound conjecture'' of \cite[Sec. 3.1]{Vafa:2005ui} where it was proposed that quantum gravity effective theories should contain at least 1 scalar field: here, apart from some peculiar examples, we propose 2. In addition, the above is consistent with (and somewhat extends) the conjecture of one systematic non-compact direction in moduli spaces \cite{Ooguri:2006in}. Note that if one wants to maintain a finite volume for the field space (see \cite{Delgado:2024skw} for a recent discussion of this idea), compact (and shrinking) field directions are in addition necessary (see e.g.~\cite[Sec. 6.1]{Palti:2019pca} for an account on this point). This hints at discussing alternatively a systematic presence of one compact and one non-compact scalar fields.

As argued above, the Field Space Statement is important for this work, and more generally for cosmology: it implies that most of the time, one should discuss scalar potentials in a multifield setting, which as explained, is very different than single field. In particular, as soon as one reaches a multifield situation, the distinction should be made between {\sl off-shell} and {\sl on-shell} statements. In addition, physical trajectories can then follow non-geodesics: this multifield possibility is known \cite{Achucarro:2018vey} to allow for new interesting cosmological evolutions.\\

\begin{figure}[t]
\begin{center}
\begin{subfigure}[H]{0.48\textwidth}
\includegraphics[width=\textwidth]{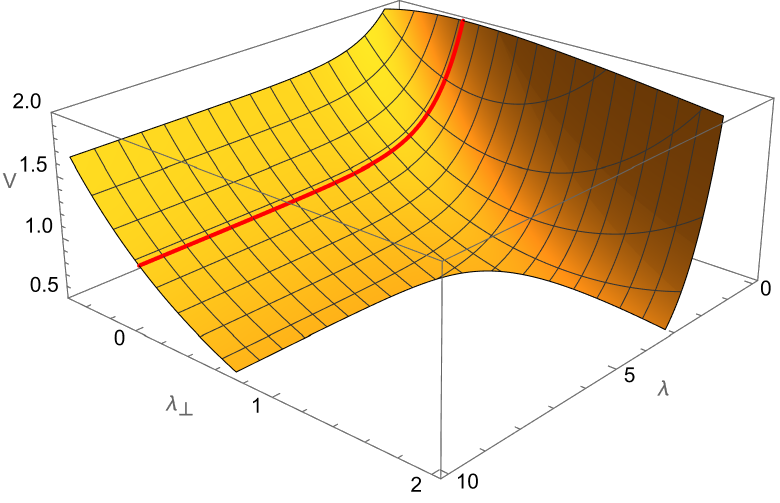}\caption{}\label{fig:Intro1}
\end{subfigure}\quad
\begin{subfigure}[H]{0.48\textwidth}
\includegraphics[width=\textwidth]{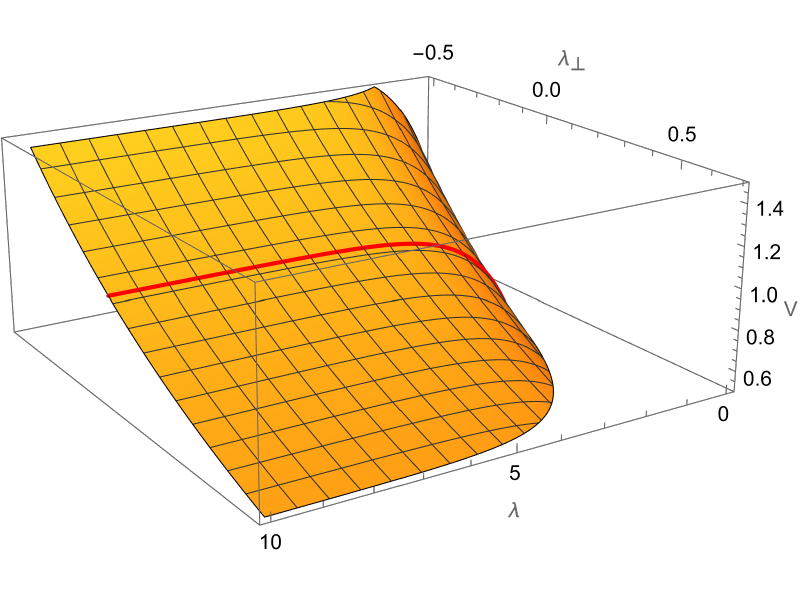}\caption{}\label{fig:Intro2}
\end{subfigure}
\caption{Potentials $V(\hat{\lambda},\hat{\lambda}_{\bot})$ allowing for an {\sl off-shell} trajectory $(\hat{\lambda},\hat{\lambda}_{\bot}=0)$, in red, such that the single slope ratio $\frac{|\del_{\hat{\lambda}} V|}{V} \rightarrow 0$ at $\hat{\lambda}\rightarrow \infty$. The general form of these potentials will be discussed in Section \ref{sec:finiteslope}, and string theory realisations will be provided in Section \ref{sec:examples}.}\label{fig:intro}
\end{center}
\end{figure}

The main result of this work is then to answer question \eqref{question1} by showing that, for $V>0$,
\bea
& \text{{\sl There exist multifield examples from string theory where in an asymptotic limit,}}\nn\\[2mm]
& \hat{\varphi} \rightarrow \infty:\qquad  \frac{|\del_{\hat{\varphi}} V|}{V} \rightarrow 0 \ .  \label{ccl}
\eea
We should specify that we {\sl do not} consider here flat directions, that would give $|\del_{\hat{\varphi}} V|=0$ everywhere in field space; those are known to exist, at least perturbatively. Here we go beyond them, finding field trajectories $(\hat{\varphi},\hat{\varphi}_{\bot}=0)$ that become flat only asymptotically along the direction considered, $\del_{\hat{\varphi}} V \rightarrow 0$, while the potential $V$ reaches a non-zero constant: see Figure \ref{fig:intro}. This general mechanism, summarized in Section \ref{sec:finiteslope}, leads to a single field slope ratio $\frac{|\del_{\hat{\varphi}} V|}{V}$ that vanishes only in the asymptotics.

We provide in Section \ref{sec:examples} and Appendix \ref{ap:extra} several examples of string effective theories in $d=4$ where the above is realised. We start in Section \ref{sec:groupmanifold} with a compactification on a group manifold. We show that trajectories realising the above can be found along potential slopes (or walls) in a $2$-dimensional field space, when $V$ diverges along one direction and goes to 0 along the other. The next examples in Section \ref{sec:CY} are within F-theory and type IIB supergravity Calabi-Yau compactifications, and a further realisation is found in Section \ref{sec:LG} with a Landau-Ginzburg model. We finally exhibit in Section \ref{sec:DGKT} a last example, found in the positive part of the potential obtained from the so-called DGKT compactification. In all cases, the limits considered correspond to field space regions where we have a good control on the corrections (e.g.~small string coupling, large volume) to the effective theory.

This allows us to conclude that the answer to question \eqref{question1} is $c=0$. In addition, we will argue in Section \ref{sec:finiteslope} that in the examples considered, by looking at a small but finite deviation of the asymptotic direction $\hat{\varphi}$, one can obtain an arbitrary value for the asymptotic single field slope ratio, within a finite range. This implies that {\sl it is meaningless to look for a non-zero lower bound to the single field slope ratio $ \frac{|\del_{\hat{\varphi}} V|}{V} $ within a multifield situation}. Insisting on {\sl off-shell} characterisations, using $\nabla V$ seems therefore better suited, as in the SdSC \eqref{SdSC}.

The considerations above ignore possible quantum gravity effects that can arise at field space asymptotics, as described by the distance conjecture or the species scale going to 0 (see e.g.~\cite{Agrawal:2019dlm, Casas:2024oak} for works studying such effects). Including those effects, especially in the case where $V$ asymptotes to a non-zero constant, could require to modify the string effective theory, therefore possibly modifying the above setting. It is hard to predict in what way our conclusions would then be altered. While these are possible loopholes, we leave these effects aside here.\\

We turn in Section \ref{sec:onshell} to the restriction to physical trajectories, motivated by question \eqref{question2} about {\sl on-shell} characterisations. Focusing on trajectories dictated by equations of motion in a cosmological background, we argue that the asymptotically flat trajectories of Figure \ref{fig:intro}, or those of Section \ref{sec:examples}, are not physical. We discuss the possibility that a physical trajectory follows, at least asymptotically, the steepest descent of a potential, and more generally a gradient flow. We argue in that case that transverse directions are flat. This reduces the gradient norm $\nabla V$, for a canonical basis, to a single field derivative along the trajectory. Still, we do not know for now of a better characterisation than the one obtained from the SdSC.

\subsection{Outlook: what about negative potentials?}

Similar questions could be asked for negative potentials: $V<0$. For such scalar potentials, analogous characterisations have been proposed \cite{Gautason:2018gln, Lust:2019zwm, Bernardo:2021wnv, Gonzalo:2021zsp, Andriot:2022bnb, Shiu:2023yzt}, including the ATCC \cite{Andriot:2022brg}. The latter gives, in a single field case, an equivalent asymptotic bound to the TCC one \eqref{TCC}: $\frac{|\del_{\hat{\varphi}} V|}{|V|} \geq \sqrt{\frac{2}{3}}$. The generalisation to multifield can then also be discussed for negative potentials. However, we have not found for $V<0$ an example of an asymptotically flat field direction where the potential would become a negative constant. For example, the DGKT compactification study in Section \ref{sec:DGKT} did not provide this, neither did the Landau-Ginzburg examples of \cite{Cremonini:2023suw}. Therefore for now, the ATCC condition suffers no counter-example even when considered in a multifield setting. It would be interesting to investigate this question further.\\

The ATCC led in \cite{Andriot:2022brg} to consider a mass bound $m^2 l^2 \leq -2$. This refers to the existence of a scalar with mass $m$ obeying this inequality, in an anti-de Sitter extremum in $d\geq 4$ with radius $l$: this property turned out to be verified in many supersymmetric examples. As a side remark, note that for $d=4$, if the bound value $\sqrt{\frac{2}{7}}$ of \cite{Calderon-Infante:2022nxb} were to be found in a negative potential, instead of the above $\sqrt{\frac{2}{3}}$, one would get $m^2 l^2 \leq -\frac{6}{7}$. Interestingly, this could accommodate most non-supersymmetric examples as well \cite[Tab. 2]{Andriot:2022brg}. More generally, if an example of an asymptotically flat direction in a negative potential was found, this could shed doubts on the claim of this mass bound. Interestingly, it is not the case. In addition, even when turning to a multifield setting, we note that the mass remains an inherently single field concept. Single field statements may then be better preserved in multifield situations for negative potentials than for positive ones. In the remainder of the paper, we focus on $V>0$.

\section{Off-shell potential characterisation in a multifield setting}\label{sec:offshell}

In this section, we first discuss the number of scalar fields to expect in an effective theory of string theory. Concluding most of the time on a multifield situation, we turn to the question \eqref{question1} on the single field slope ratio in the asymptotics. We summarise the mechanism that gives a vanishing bound $c=0$, or an arbitrary value within a finite range. This mechanism will be exemplified in Section \ref{sec:examples}.

\subsection{How many fields?}\label{sec:singlefield}

As discussed in the Introduction, models of the type \eqref{action} that are truly single field avoid certain ambiguities: the notion of asymptotic is clearly defined and can typically be reached via an {\sl on-shell} trajectory (unless the field is stuck in a vacuum). Can one get such a single field model as a string effective theory?

When considering a standard geometric compactification to a $d$-dimensional effective theory, $3\leq d \leq 9$, one necessarily faces at least two scalar fields: the dilaton and the volume of the compact space. In particular, both the dilaton and the metric are even fields under an orientifold involution so the resulting scalar fields are not projected out. Note also that they are ``non-compact'', in the sense that their allowed values span an infinite range, and they are neutral, meaning they are not charged under a gauge group. To get a truly single field model, one should then either not compactify (i.e.~stay in $d=10$ string theory, or even $d=11$ where M-theory admits no scalar field at all), or get to a lower dimensional effective theory without a standard compactification (e.g.~via non-geometric constructions). Let us discuss these two options, and connect to the {\bf Field Space Statement} proposed and discussed around \eqref{conj}.

10-dimensional string theories admit the dilaton scalar field in their spectrum on a Minkowski background. At least in a perturbative setting, we do not know of a mechanism that would remove the dilaton as a relevant degree of freedom in an effective theory; this gives one scalar field. Considering the tree-level approximation, we know of several ways to generate a scalar potential; let us examine those. One may first consider 10d type IIA supergravity with Romans mass, as discussed in \cite[Sec. 4.2]{Andriot:2023wvg}: this is the 10d analogue of a scalar potential generated by an ``internal'' flux (meaning on the would-be compact space). With a canonically normalized dilaton, the potential was shown to be an exponential with rate $\frac{5}{\sqrt{2}}$; note that this is greater than the SdSC value $2/\sqrt{d-2}$ for $d=10$. Therefore, even with a truly single field model, the single field slope ratio is not particularly low. In type IIB, one may also consider $O_9/D_9$ sources. One can imagine getting a potential for the dilaton via the DBI action.\footnote{\label{foot:O9}Because of their trivial Bianchi identity, $O_9/D_9$ have a priori to provide an opposite contribution to the potential, in such a way that actually, no potential term is eventually generated for their common volume or for the dilaton; see \cite[(2.3)]{Andriot:2022xjh}.} The term goes as $e^{\frac{3}{2} \phi}$, which in 10d amounts to $\tau^{-6}$  (see e.g.~\cite[(4.17),(4.14),(4.6)]{Andriot:2022xjh}). Once canonically normalized, this gives an exponential potential with rate $\frac{3}{\sqrt{2}}$. The latter is once again greater than $2/\sqrt{d-2}$.

Let us now comment on M-theory and 11-dimensional supergravity. The latter admits no scalar field. As a consequence, a suitably chosen compactification manifold could leave the volume as the only scalar field in a lower dimensional effective theory, contrary to the situation described above for string theory.\footnote{We thank Niccolo Cribiori for related discussions.} For instance, the very specific choice of a rigid Calabi-Yau three-fold towards a $d=5$ theory, or an analogous 7-dimensional $G_2$-manifold towards $d=4$, would give a single field effective theory without potential. Considering a 7-dimensional hyperbolic compact manifold would also leave just the volume, this time with a potential generated by curvature. Such an example was considered in \cite{Andersson:2006du}, the potential being analogous to \eqref{potcurv} but from an 11-dimensional origin. These peculiar examples indicate that an extension of the {\bf Field Space Statement} from string theory, as written for now, to M-theory, and beyond this to quantum gravity, should be done with care. In addition, the manifolds required to have only one field in lower dimensions, $d\leq 9$, are such that their geometry does not allow to single out a circle. This implies that these examples precisely do not admit a perturbative limit where one would recover (type IIA) string theory. Insisting on viewing them as stringy examples, they would correspond to non-perturvative, strongly coupled and isolated string examples. Asymmetric orbifold examples below, giving only one non-compact scalar field, will be reminiscent of these features.

Turning to the possibility of having a lower dimensional effective theory without a volume field, one may first consider Landau-Ginzburg models \cite{Greene:1996cy, Becker:2006ks}. Those however have complex structure moduli, in addition to the (axio-)dilaton: we will consider examples of those in Section \ref{sec:examples}. Imaginary parts of the complex structure moduli are also non-compact scalars. Would it then be possible for Landau-Ginzburg models to have no complex structure moduli? This would mean having a world-sheet superpotential without marginal deformation, which seems difficult to realise. In particular, restricting ourselves to symmetric Gepner models \cite{Gepner:1987qi}, there is no example with $h^{1,1}=h^{2,1}=0$. Similarly, one could consider hypothetically an F-theory setting on a ``Calabi-Yau fourfold'' mirror dual to a rigid one (i.e.~without K\"ahler modulus). It seems once again difficult to have an elliptic one, with only one complex structure modulus (that would correspond to the axio-dilaton). Finally, one may also investigate ``non-geometric'' settings in the sense of T- or S-folds \cite{Plauschinn:2018wbo}: at first sight, those do not offer less fields either.

More non-geometric constructions, with exotic effective theories, appear when considering asymmetric orbifolds. The work \cite{Harvey:1987da} presents a $d=2$ example with no scalar field at all! More relevant to our setting, \cite{Dabholkar:1998kv} gives examples in $d=4$ with only 1 non-compact scalar, the dilaton, provided there are 16 or more supercharges; further examples were recently provided in \cite{Aldazabal:2025zht}. This suggests the need to consider a lower number of supersymmetries, in order to have more non-compact scalar fields. However, recent works provide similar examples with less supersymmetries. To start with, references and new examples of $d=6$ constructions with 8 supercharges and no neutral hypermultiplet are given in \cite{Baykara:2023plc}. Such examples leave only ($d=6$) tensor multiplets to provide one non-compact scalar,\footnote{A list of supermultiplets in even dimensions, and their content according to supersymmetry, can be found e.g.~in \cite[App. A]{Blumenhagen:2016rof}.} and among them, one finds in \cite{Baykara:2023plc, Baykara:2024vss} instances with only one tensor multiplet. Reducing further supersymmetry to 4 supercharges, one $d=4$ example, the heterotic $\mathbb{Z}_{24}$ orbifold given in \cite[Tab. 10]{Baykara:2024vss}, has only one non-compact neutral scalar in its single neutral matter multiplet. Finally, \cite{Baykara:2024tjr} exhibits examples of non-supersymmetric string constructions to $d=4$ with again only one neutral scalar. Regarding those last examples, even though they are tachyon-free, the absence of supersymmetry makes the control on corrections a priori poorer; we refer to this work for related discussions. Still, we see for now that these less conventional ways of obtaining lower dimensional effective theories of string theory provide few, very specific examples with only one non-compact neutral scalar. Those counter-examples to a complete claim of 2 non-compact scalars remain for now isolated, especially in $d=4$ with ${\cal N} \leq 1$, relevant to phenomenology.

As a consequence, we claim that in most examples known, for $3\leq d \leq 9$, a $d$-dimensional string effective theory has a scalar field space with dimension at least 2, meaning that it is {\sl multifield}. Those 2 scalars do not include axions: they are along non-compact field directions. This is all in line with the {\bf Field Space Statement} proposed in \eqref{conj}. A multifield situation opens the possibility of {\sl off-shell} considerations, that will be the topic of the next sections.\\

Let us now come back to the standard dilaton and (internal) volume field, and focus on a $d=4$ effective theory. Is it then possible to have only those two fields? To that end, one should first avoid having other ``geometric moduli'' or volume fields of internal subspaces. This is possible if for example, we take a rigid Calabi-Yau, or a hyperbolic space to get curvature (divided by a lattice to have it compact). With such a 6d compact space, it is difficult to have (1- to 5-form) fluxes, or $O_p/D_p$ wrapping subspaces;\footnote{The argument may not hold for $p=9$, for which the wrapped subspace is the whole internal space, so there is no extra volume field introduced. However, as explained in Footnote \ref{foot:O9}, a potential term for the volume cannot be generated in this way.} otherwise such a subspace would lead to an extra (volume) field. In a type IIA supergravity compactification, such a specific 6d geometry avoids for the same reason to get axions: we are then indeed left with only the dilaton and the 6d volume.

What is then the scalar potential? In type IIA compactifications, one could still consider the Romans mass, as well as the constant 6-form flux (10d Hodge dual of a spacetime filling $F_4$-flux): for simplicity, let us consider a configuration without them. The main contributor to the (perturbative, leading order) scalar potential is then the internal curvature term, given more generally in $d$ dimensions by
\beq
2 V = e^{- \left( \frac{2}{\sqrt{d-2}}\, \hat{\tau} + \sqrt{\frac{4}{10-d}}\, \hat{\rho} \right) } \left( - {\cal R}_{10-d} \right) \ ,\nn
\eeq
in terms of the canonically normalised $d$-dimensional dilaton $\hat{\tau}$ and the volume $\hat{\rho}$ for $3 \leq d\leq 9$ \cite[(5.3)]{Andriot:2022brg}. We can introduce a new canonical basis $(\hat{\lambda},\hat{\lambda}_{\bot})$
\beq
\alpha\, \hat{\lambda} = \frac{2}{\sqrt{d-2}}\, \hat{\tau} + \sqrt{\frac{4}{10-d}} \, \hat{\rho} \ ,\quad
\alpha\, \hat{\lambda}_{\bot} = - \sqrt{\frac{4}{10-d}}  \,\hat{\tau} + \frac{2}{\sqrt{d-2}} \, \hat{\rho} \ , \quad \alpha= \frac{4 \sqrt{2}}{\sqrt{(d-2)(10-d)}} \nn
\eeq
such that the potential gets rewritten as
\beq
2 V = e^{- \frac{4 \sqrt{2}}{\sqrt{(d-2)(10-d)}} \,  \hat{\lambda} } \left( - {\cal R}_{10-d} \right) \ .\label{potcurv}
\eeq
Note that the corresponding rate is greater than the SdSC one thanks to $d>2$, so such a restrictive geometry to limit the number of fields does not help in that respect.\footnote{A similar compactification to $d=4$ is discussed in \cite[Sec. 4.1.1]{Andriot:2023wvg} with the same rate, namely $\frac{4}{\sqrt{6}}$, where it is however claimed that only the internal volume appears in the potential. This is because the 6d curvature and associated volume field is there defined in 10d Einstein frame, and therefore absorbs the dilaton field, while in the potential above, from \cite[(5.3)]{Andriot:2022brg}, the curvature is defined in 10d string frame and we thus got a dependence on the dilaton. The above field redefinition allows to go from one to the other, by redefining what one calls the internal volume field.} Interestingly, in that example, we end-up with only one field in the potential and one flat direction: we could call this situation an {\sl almost single field} example. These observations are based on this classical (perturbative) potential; it would be interesting to see whether higher corrections could alter the conclusion on the rate, and generate potential terms for the second field.\\

Finally, for completeness, let us mention that one may consider an effectively single field model: this refers to having e.g.~all fields stabilised except one which is rolling, so that the dynamics is determined by a single field. Such a situation is however only of interest for {\sl on-shell} considerations, so we will come back to it in Section \ref{sec:onshell}. To allow for stabilisation, one would need more terms in the potential than considered above, which as explained, could be achieved thanks to a less rigid manifold, so at the cost of introducing more fields, going further away from a truly single field model.

\subsection{The bound on the single field slope ratio}\label{sec:finiteslope}

We have argued that $d$-dimensional string effective theories for $3\leq d \leq 9$ are most of the time multifield. This allows one to consider {\sl off-shell} trajectories, meaning $1$-dimensional subspaces of the field space that are not physical trajectories, and study single field slope ratios $\frac{|\del_{\hat{\lambda}} V|}{V} $ along them. Let us consider the case of two (canonically normalized) fields $(\hat{\lambda}, \hat{\lambda}_{\bot})$, where the trajectory and asymptotic of interest are given by $\hat{\lambda}_{\bot}=0,\ \hat{\lambda} \rightarrow \infty$. Furthermore, we take a scalar potential of the following form in this asymptotic
\beq
V(\hat{\lambda}, \hat{\lambda}_{\bot}) = a_1 \, e^{c_1\, \hat{\lambda}_{\bot}} + a_2 \, e^{c_2\, \hat{\lambda}_{\bot} + b_2\, \hat{\lambda}} + o(e^{b_2\, \hat{\lambda}}) \,,
\quad b_2<0 \,,
\quad
a_1 a_2 \neq 0 \,,
\quad c_1 \neq 0\ . \label{Vgen}
\eeq
Illustrations of such a potential can be found in Figure \ref{fig:intro}: we took $V= e^{- \hat{\lambda}_{\bot}} + e^{\hat{\lambda}_{\bot} - \hat{\lambda}}$ in Figure \ref{fig:Intro1}, and $V= e^{- \hat{\lambda}_{\bot}} - e^{\hat{\lambda}_{\bot} - \hat{\lambda}}$ in Figure \ref{fig:Intro2}. As can be seen there, one gets a vanishing single field slope ratio in the asymptotic along the trajectory of interest. Indeed, one obtains generically from the potential \eqref{Vgen}
\begin{align}
\hat{\lambda}_{\bot}=0\,,
\quad
\hat{\lambda} \rightarrow \infty\,:
&
\qquad \del_{\hat{\lambda}} V \rightarrow 0 \ ,\  V \rightarrow {\rm constant} > 0 \,, \label{genmech}\\
\Rightarrow
&
\quad\quad
\frac{|\del_{\hat{\lambda}} V|}{V} \rightarrow 0\, .
\end{align}
More generally, a sufficient (but not necessary) condition on the potential for this mechanism \eqref{genmech} to be realized is to have a trajectory $\hat{\lambda}_{\bot}^1=\hat{\lambda}_{\bot}^2= \dots =0,\,\hat{\lambda} \rightarrow \infty$, such that the potential takes the form
\beq
V(\hat{\lambda}, \hat{\lambda}_{\bot}^i) = V_1 (\hat{\lambda}_{\bot}^i) + V_2(\hat{\lambda}, \hat{\lambda}_{\bot}^i)\ ,\qquad {\rm with}\ \ V_1(0)>0 \ ,\ \ V_2\,(\hat{\lambda}, 0) \rightarrow 0 \ ,\ \  \del_{\hat{\lambda}} V_2\,(\hat{\lambda}, 0) \rightarrow 0 \ .
\eeq
In Section \ref{sec:examples}, we will present several string theory examples of the above situation. The general mechanism \eqref{genmech}, giving a vanishing single field slope ratio in the asymptotic, answers the question \eqref{question1} by providing a lower bound $c=0$.\\

Let us now consider a small deviation from the $\hat{\lambda}$-trajectory discussed around \eqref{Vgen}: one expects the new trajectory to provide a small non-zero single field slope ratio. To show this, we perform a field rotation to another canonical field basis $(\hat{\lambda}, \hat{\lambda}_{\bot}) \rightarrow (\hat{\eta}, \hat{\eta}_{\bot})$
\begin{equation}
\begin{split}
\hat{\lambda} &= c_{\epsilon}\, \hat{\eta} - s_{\epsilon} \, \hat{\eta}_{\bot} \\
\hat{\lambda}_{\bot} &= s_{\epsilon}\, \hat{\eta} + c_{\epsilon}\, \hat{\eta}_{\bot} \\
{\rm where}
\quad\quad
c_{\epsilon} &\equiv \cos(\epsilon) \,,
\quad s_{\epsilon} \equiv \sin(\epsilon) \ .
\end{split}
\end{equation}
The potential becomes
\beq
V(\hat{\eta}, \hat{\eta}_{\bot}) = a_1 \, e^{c_1 s_{\epsilon}\, \hat{\eta}} \, e^{ c_1 c_{\epsilon} \, \hat{\eta}_{\bot}} + a_2 \, e^{ (c_2 s_{\epsilon} + b_2 c_{\epsilon})\, \hat{\eta}}  \, e^{ (c_2 c_{\epsilon} - b_2 s_{\epsilon})\, \hat{\eta}_{\bot}} + o(...) \ ,
\eeq
where we will comment on the neglected terms $o(...) $. We now consider a different trajectory and asymptotic
\beq
\hat{\eta}_{\bot}=0 \,,
\quad\quad
\hat{\eta} \rightarrow \infty \ .
\eeq
Essentially, we rotated the trajectory and now follow the new one which has a non-zero slope. Still, for a small angle $\epsilon$, the new slope is expected to be small, since the two trajectories are expected to be ``close'' to each other. This can be estimated as follows: for a small enough $\epsilon$ (one can imagine $s_{\epsilon} \sim 0\ ,\ c_{\epsilon} \sim 1$, although these quantities actually do not have to be infinitesimal), one has
\beq
c_2 s_{\epsilon} + b_2 c_{\epsilon} <0
\quad
{\rm and}
\quad
|c_1 s_{\epsilon}| < |c_2 s_{\epsilon} + b_2 c_{\epsilon}| \ ,
\eeq
which ensures that the second term in the potential is negligible compared to the first one along the $\hat{\eta}$-trajectory and asymptotic, given by $\hat{\eta}_{\bot}=0 \ ,\ \hat{\eta} \rightarrow \infty$. The above conditions give a finite (not infinitesimal) upper bound to the angle. Turning to the neglected terms $o(...)$, those were initially dependent on $(\hat{\lambda}, \hat{\lambda}_{\bot})$, and negligible compared to the second term in the potential for the $\hat{\lambda}$-trajectory. Their $\hat{\lambda}$ dependence now becomes $c_{\epsilon} \hat{\eta}$ along the $\hat{\eta}$-trajectory ($\hat{\eta}_{\bot}=0$), and their $\hat{\lambda}_{\bot}$ dependence becomes $s_{\epsilon} \hat{\eta}$. As above, for a small enough $\epsilon$ (i.e.~provided a certain upper bound), these terms remain negligible: their behaviour will essentially be given by their dependence in $c_{\epsilon} \hat{\eta}$, which will be negligible compared to the second potential term. In addition, since the potentials considered only have a finite number of terms, we end up with a small but finite overall upper bound $\epsilon_0$, that allows to keep the same hierarchy in the potential terms between the two trajectories. This implies
\begin{align}
|\epsilon| < \epsilon_0:\quad \hat{\eta}_{\bot}=0 \ ,\ \hat{\eta} \rightarrow \infty
\,\,:
&\quad\quad
V(\hat{\eta}, \hat{\eta}_{\bot}) = a_1 \, e^{c_1 s_{\epsilon}\, \hat{\eta}} + o(e^{c_1 s_{\epsilon}\, \hat{\eta}})\\
\Rightarrow
&\quad\quad
\frac{\del_{\hat{\eta}} V}{V} \rightarrow c_1 s_{\epsilon} \, .
\end{align}
We have shown that provided a $\hat{\lambda}$-trajectory where the single field slope ratio vanishes in the asymptotic, it is easy to build an infinite set of trajectories which have a single field slope ratio ranging between $0$ and a finite, non-infinitesimal, value, given here by $|c_1| \sin \epsilon_0$. In addition to having  examples with a vanishing single field slope ratio in the asymptotic, this proves that {\sl there exists no non-zero lower bound to this ratio.}\\

We conclude that the single field slope ratio is not suited to an {\sl off-shell} potential characterisation in a (multifield) string effective theory, and that $\nabla V$ is a better quantity, as e.g.~in the SdSC \eqref{SdSC}. We provide in the next section string theory examples of the above mechanism, that led us to this conclusion.

\section{String theory off-shell examples}\label{sec:examples}

We present in this section several examples of 4d theories of the form \eqref{action}, derived from string theory, that realise the mechanism \eqref{genmech} discussed in Section \ref{sec:finiteslope}. This means that in each example, we identify an {\sl off-shell} trajectory along $(\hat{\lambda}, \ \hat{\lambda}_{\bot}=0)$ with an asymptotic direction $\hat{\lambda} \rightarrow \infty$, such that in this limit, one gets a vanishing single field slope ratio $\frac{|\del_{\hat{\lambda}} V|}{V} \rightarrow 0$ (with $V>0$).

In the examples to be presented, the asymptotic considered corresponds to a limit where string theory corrections to the 4d theory are under control (e.g.~weak string coupling, large volume), hence the claim of ``string theory examples''.\footnote{We would like to point out to the reader that in setups involving smeared sources, it remains debatable whether the full solutions with localized sources would preserve their properties or show significant changes, and whether they can confidently be considered string theory solutions rather than supergravity solutions. For further discussion, we refer to the following papers \cite{Blaback:2010sj, Junghans:2020acz, Marchesano:2020qvg, Baines:2020dmu, Cribiori:2021djm ,  Emelin:2022cac, Emelin:2024vug, VanHemelryck:2024bas}, which explore the reliability and the process of unsmearing in such solutions, and to \cite{McOrist:2012yc, Bardzell:2024anh} which address both earlier and more recent concerns.} Strictly speaking, for some examples, one should also verify that the point picked in the transverse hyperplane, namely the values taken by the transverse fields to define the trajectory (see footnote \ref{foot:asymptotic}), are consistent with such a control; we briefly discuss this matter.

We start in Section \ref{sec:groupmanifold} with a compactification of 10d type IIA supergravity on a group manifold: the resulting 4d theory provides an explicit realisation of the schematic potential \eqref{Vgen} and the mechanism \eqref{genmech}; the next examples give generalisations of those, with interesting features. We will consider in Section \ref{sec:CY} an F-theory Calabi-Yau fourfold (CY${}_4$) and a type IIB CY${}_3$ compactifications, followed in Section \ref{sec:LG} by a Landau-Ginzburg model effective theory. The last example to be presented in Section \ref{sec:DGKT} is a little different: this so-called DGKT compactification gives rise to a scalar potential famous for its negative potential minimum, an anti-de Sitter solution. We will show that this 4d potential becomes positive in an asymptotic direction of interest.

\subsection{Group manifold compactification}\label{sec:groupmanifold}

Starting with 10d type II supergravities, taken as low energy and classical string effective theories, one can consider a compactification on a 6d group manifold, together with fluxes and orientifolds, leading by dimensional reduction to a 4d theory. In such a setting, one can find 10d solutions with a maximally symmetric 4d spacetime. A classification and analysis of such solutions was provided in \cite{Andriot:2022way, Andriot:2022yyj}. Furthermore, these compactifications on group manifolds allow for a specific dimensional reduction, a consistent truncation, giving a 4d theory of the form \eqref{action}: this specificity implies that one can look for extrema of the 4d potential, and those will correspond to 10d solutions with maximally symmetric 4d spacetimes. This 4d-10d interplay can be very useful, and this specific dimensional reduction (consistent truncation) was studied in detail and automatized in \cite{Andriot:2022bnb} for the complete classification of compactifications considered in \cite{Andriot:2022way, Andriot:2022yyj}. In other words, thanks to the code {\tt MSSV} of \cite{Andriot:2022bnb}, one systematically obtains a 4d theory of the form \eqref{action}, namely the scalar fields, their kinetic terms and potential, for each class of compactification on group manifolds. And for some of those, examples of solutions (extrema of the potential) are known. We are going to use such an example in the following.

To get a positive scalar potential in that context, a starting point would be an example with a de Sitter solution: at least, the 4d potential around the corresponding extremum should be positive. One may discuss whether this 4d theory can be trusted as an effective string theory. This question can be asked already at the de Sitter solution point, and this is actually a non-trivial question, discussed recently in \cite{Andriot:2024cct}. However, as mentioned above, we will focus here on an asymptotic where (at least some) string corrections are under control, giving confidence on the 4d theory, in that field space region of interest.\\

We consider here the class of type IIA supergravity compactifications denoted $s_{6666}$: it can be understood as having four intersecting sets of $O_6$-planes, preserving ${\cal N}=1$ supersymmetry in 4d. The latter might be broken by background fluxes additionally turned on. In this class, the 4d theory admits 14 scalar fields: 7 axions (coming from the 10d fields $B_2$ and $C_3$) and 7 saxions (corresponding to 6 diagonal metric fluctuations $g_{aa}$ and the dilaton $\phi$). The field space metric is block diagonal between those two sectors. This allows us in the following to ignore the axions: in practice, we set them to 0, which is the value they conventionally take at a potential extremum. We recall that giving a value to transverse fields as here is part of defining an ({\sl off-shell}) field space trajectory (see footnote \ref{foot:asymptotic}). Turning to the saxions, we can read their kinetic terms from {\tt MSSV}. They can be recast in the familiar canonical form
\begin{equation}
\begin{split}
\frac{1}{2} g_{ij} \del_{\mu} \varphi^i \del^{\mu} \varphi^j
&= \frac{1}{8} \sum_{a=1}^6 (\del \ln g_{aa})^2 +  (\del \ln \tau)^2 \,,
\quad {\rm where} \quad
\tau= e^{-\phi} \left(\prod_a g_{aa}\right)^{\frac{1}{4}} \\
&= \frac{1}{2} \sum_{a=1}^6 (\del \hat{g}_{aa})^2 +  \frac{1}{2} (\del \hat{\tau})^2\,,
\quad\quad\,\,\,\,{\rm where} \quad
\hat{g}_{aa} = \frac{1}{2} \ln g_{aa} \,,
\quad \hat{\tau} = \sqrt{2} \ln \tau \ .
\end{split}
\end{equation}
This simplicity is due to the underlying 4d ${\cal N}=1$ structure of this type IIA example.

The scalar potential depends on the details of the background fluxes and of the compact group manifold chosen. We choose here those specifics in such a way that a de Sitter solution exists, as motivated above to get a positive scalar potential. In other words, we take the de Sitter solution $s_{6666}^+ 4$ \cite{Andriot:2022way, Andriot:2022yyj}, which fixes all background quantities, including the group manifold, and we consider the 4d theory around and away from this critical point of $V$.

Let us mention that this extremum admits one tachyonic field direction. Asymptotics of de Sitter tachyonic directions were discussed in \cite{Andriot:2023isc}, including the present example, while looking for such directions along which the scalar potential would asymptotically vanish.\footnote{In \cite{Andriot:2023isc}, the scalar potential of the solution $s_{6666}^+ 4$ was studied along the tachyonic direction $y$, and it could be seen to exhibit a ``wall'' at large $y$, i.e.~a divergence of the potential. A more careful study indicates that this is due to $g_{22}\rightarrow 0$. Unfortunately for the present work, this asymptotic does not allow for a good control on the supergravity approximation, so we cannot make use of this ``wall''. Another ``wall'' appears in the opposite $y$ direction, which leads to a large $e^{\phi}$, where we face the same issue. So we consider here a different direction.} But we will consider here a different asymptotic direction. Turning to flat directions of the potential, the 4d theory obtained in this compactification class generically admits none. However, our specific compactification example with a de Sitter solution does: one flat direction appears, as a combination of $C_3$ axions, $C_{3\, 346}-C_{3\, 145}$. Restricting to the saxions in the following, we can be sure that the field directions to be considered do not involve a flat direction.\\

\begin{figure}[t]
\begin{center}
\begin{subfigure}[H]{0.48\textwidth}
\includegraphics[width=\textwidth]{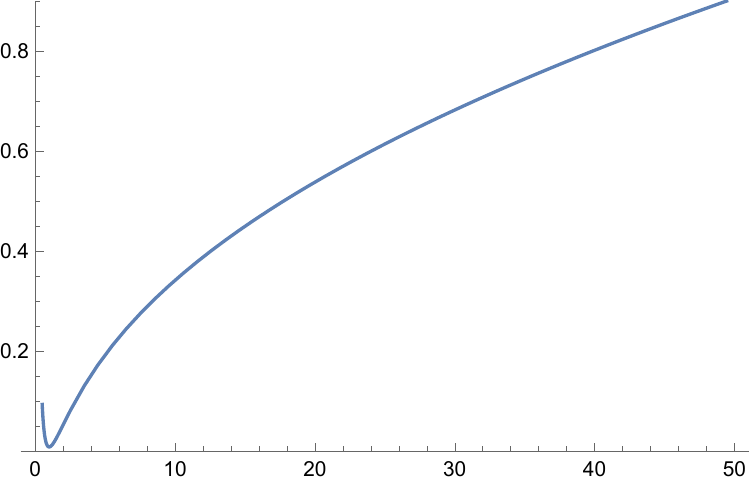}\caption{$V(g_{11})$}\label{fig:Vg11}
\end{subfigure}\quad
\begin{subfigure}[H]{0.48\textwidth}
\includegraphics[width=\textwidth]{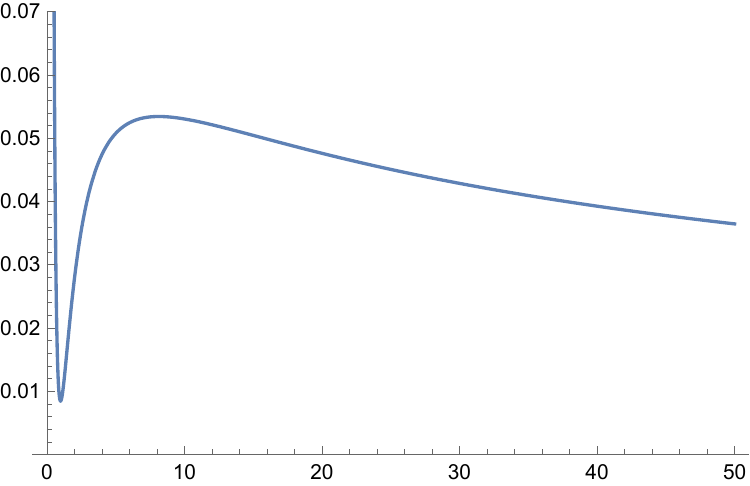}\caption{$V(g_{44})$}\label{fig:Vg44}
\end{subfigure}\\
\begin{subfigure}[H]{0.48\textwidth}
\includegraphics[width=\textwidth]{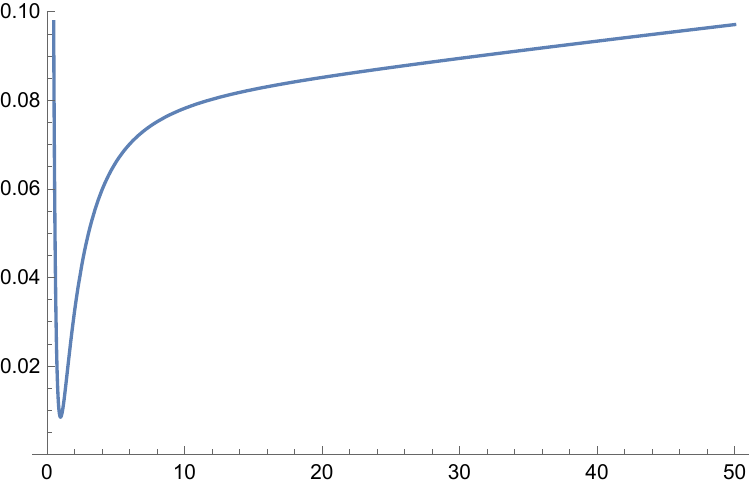}\caption{$V(g_{55})$}\label{fig:Vg55}
\end{subfigure}\quad
\begin{subfigure}[H]{0.48\textwidth}
\includegraphics[width=\textwidth]{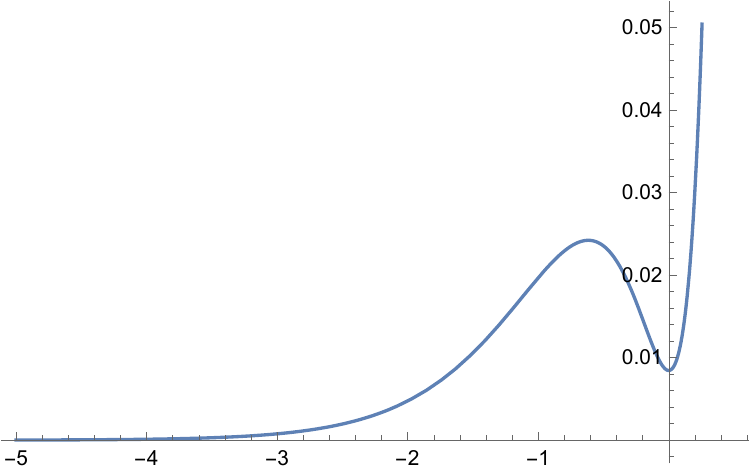}\caption{$V({\phi})$}\label{fig:Vp}
\end{subfigure}
\caption{Scalar potential $V$ obtained with the de Sitter solution $s_{6666}^+ 4$, setting all axions and saxions to their critical point value except for one saxion: this gives the potential profile along that saxionic direction. The profile for $g_{11}$ and $g_{22}, g_{33}$ are very similar, those for $g_{55}$ and $g_{66}$ as well.} \label{fig:Vs66664}
\end{center}
\end{figure}

To find the field direction of interest, let us depict the profile of the scalar potential for the various saxions in Figure \ref{fig:Vs66664}.

We see in Figure \ref{fig:Vs66664} that $g_{11},g_{22},g_{33},g_{55},g_{66}$ appear as stabilised along their own direction, with a divergent positive potential in the classical direction, namely for large $g_{aa}$, where we get corrections under control. On the contrary, $g_{44}$ and $\phi$ display a metastable profile, with the positive potential going to 0 asymptotically in their classical direction (large $g_{44}$ and small $e^{\phi}$). Note that the observed stability is only apparent, since instability can occur in this sector due to off-diagonal terms in the Hessian (see \cite[Sec.3.3]{Andriot:2020wpp} and \cite[Fig.7]{Andriot:2023isc}).\footnote{It is actually the case here: we know that a tachyon appears in a subsector of 5 saxions \cite{Andriot:2022yyj}, and we also know that the tachyon observed within the 14 fields has its main contributions from the saxions \cite{Andriot:2022bnb}.} What matters to us here is however not the stability at the critical point, but rather the asymptotic in the classical direction. Indeed, if one direction goes asymptotically to 0 and another one goes to $+\infty$, both towards a region where the potential is controlled, then the potential can exhibit a slope or even a wall in this 2d field space, all the way to a trustable asymptotic. A potential with a slope or a wall should admit a field direction which behaves asymptotically as desired: it becomes flat. Building on this idea, we will find the appropriate field directions to realise the mechanism \eqref{genmech}. For example, from {\tt MSSV}, we obtain the following dominant term of the potential in a classical asymptotic
\beq
V(g_{11},g_{44},\phi) \sim 0.13351\, e^{2\phi } \, \sqrt{\frac{g_{11} }{g_{44}^3}}
\quad\quad {\rm for}\quad
g_{11},g_{44}, e^{-\phi} \sim +\infty \ , \label{Vdom14p}
\eeq
where the other fields have been set to their critical point values ($g_{aa}=1$, where the fields $g_{aa}$ should be understood as fluctuations around central background values). From this expression, it is easy to find combinations of fields, or curves, which give an asymptotically constant non-zero potential.\\

\begin{figure}[H]
\centering
\includegraphics[width=0.6\textwidth]{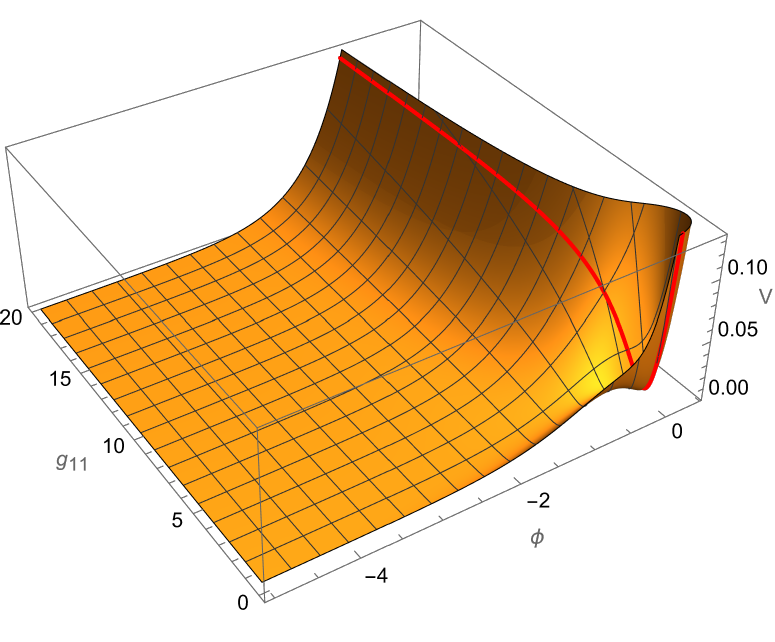}
\caption{$V(\phi,g_{11})$ (the other fields being set to their critical point value), with the curve $g_{11} = e^{-4\phi}$ (in red), which provides an asymptotically flat direction.}\label{fig:V1pcurve}
\end{figure}

For simplicity, we further set $g_{44}=1$, and we give the following two-field potential, with the first subdominant term in the desired asymptotic
\beq
V(\phi,g_{11}) \sim  0.13351\, {g_{11}}^{\frac{1}{2}}\, e^{2\phi} + 0.16790\, {g_{11}}^{-\frac{1}{2}}\, e^{2\phi} \quad\quad {\rm for}\quad
g_{11},e^{-\phi} \sim +\infty \ .\label{Vdom1p}
\eeq
This potential goes asymptotically to a constant with the curve
\beq
g_{11} = e^{-4\phi} \ \Rightarrow \ V(\phi,g_{11})   \rightarrow 0.13351 \ .
\eeq
We depict this curve in Figure \ref{fig:V1pcurve}.
In terms of canonical fields, the limit of interest becomes $\hat{g}_{11},\hat{\tau} \rightarrow +\infty$, and the two-field potential \eqref{Vdom1p} gets rewritten in this limit as
\beq
V(\hat{g}_{11},\hat{\tau}) \sim 0.13351\, \tau^{-2} \, g_{11} +  0.16790\, \tau^{-2}   =  0.13351\, e^{-\sqrt{2} \, \hat{\tau} + 2 \, \hat{g}_{11}}  +  0.16790\, e^{-\sqrt{2} \, \hat{\tau}} \ .
\eeq
The curve of interest is now given by $ \hat{\tau} = \sqrt{2} \, \hat{g}_{11} $, which obviously sets the potential asymptotically to a constant.

We finally introduce an appropriate pair of canonically normalised fields, $(\hat{\lambda}, \hat{\lambda}_{\bot})$, as follows
\beq
\hat{\lambda}= \sqrt{\frac{2}{3}}\, \hat{\tau} + \frac{1}{\sqrt{3}} \, \hat{g}_{11} \,,
\quad\quad
\hat{\lambda}_{\bot}= -\frac{1}{\sqrt{3}} \, \hat{\tau} +  \sqrt{\frac{2}{3}}\, \hat{g}_{11} \ ,
\eeq
where one can verify that $(\del \hat{\lambda})^2 + (\del \hat{\lambda}_{\bot})^2 = (\del \hat{\tau})^2 +(\del \hat{g}_{11})^2$. The curve, or {\sl off-shell} trajectory, is now along $\hat{\lambda}$ with $\hat{\lambda}_{\bot} = 0$, and the asymptotic of interest is $\hat{\lambda} \rightarrow + \infty$. In this limit, the potential gets rewritten as
\beq
V(\hat{\lambda} ,\hat{\lambda}_{\bot}) \sim  0.13351\, e^{-\sqrt{6} \, \hat{\lambda}_{\bot}} +  0.16790\, e^{\sqrt{\frac{2}{3}} \, \hat{\lambda}_{\bot} \,  -\frac{2}{\sqrt{3}} \, \hat{\lambda} } + o(e^{-\frac{2}{\sqrt{3}} \, \hat{\lambda} }) \ ,
\eeq
i.e.~in the exact same form as \eqref{Vgen}. This allows to see the general mechanism \eqref{genmech} realised, namely
\beq
\hat{\lambda}_{\bot} = 0 \ ,\ \hat{\lambda} \rightarrow + \infty : \quad \del_{\hat{\lambda}} V \rightarrow 0 \ ,\  V \rightarrow {\rm constant} > 0 \ .
\eeq
This shows explicitly that the single field slope ratio is vanishing in this asymptotic.\\

Having presented in detail the idea and the formalism to find a field direction realising the mechanism \eqref{genmech}, we now turn to more straightforward examples, which generalise the latter.

\subsection{Calabi-Yau compactifications}\label{sec:CY}

\subsubsection{\texorpdfstring{F-theory and CY\textsubscript{4}}{}}\label{sec:CY4}

We start with an example of F-theory compactifications on CY${}_4$; the resulting 4d theories, especially the scalar potentials, were recently studied in \cite{Calderon-Infante:2022nxb} that we will follow. The K\"ahler moduli were not discussed in \cite{Calderon-Infante:2022nxb}, and without further ingredients, they would not be stabilized but would roll-down the potential (see below). Discarding them, or equivalently setting them to a given value, amounts to consider {\sl off-shell} field trajectories. We do so here, focusing on the complex structure moduli space.

In \cite{Calderon-Infante:2022nxb} were considered CY${}_4$ with 2 complex structure moduli, that we denote here $\tau,U$; the map between our conventions and those of \cite{Calderon-Infante:2022nxb} can be found in Appendix \ref{ap:CY}. We restrict to Sen's weak string coupling limit, where the setup reduces to a type IIB orientifold compactification, $\tau$ corresponds to the 10d axio-dilaton and $\text{Im}\,\tau \rightarrow \infty$. We also consider the large complex structure limit where $\text{Im}\,U \rightarrow \infty$. Following \cite{Grimm:2019ixq, Calderon-Infante:2022nxb}, we then write the K\"ahler potential in this limit as
\be\label{eqn:Kahlerpotential}
    K = - \log[-\mathrm{i}(\tau - \bar{\tau})] - 3 \log[\mathrm{i}(U-\bar{U})]\, ,
\ee
where we do not consider the K\"ahler moduli contribution. The kinetic terms, expressed with the K\"ahler metric, are then given by
\be \label{kin}
K_{I \bar{J}} \partial_{\mu}\psi^{I}\partial^{\mu}\bar{\psi}^{\bar{J}} = -\frac{1}{(\tau-\bar{\tau})^2} \partial_{\mu} \tau \partial^{\mu} \bar{\tau}-\frac{3}{(U-\bar{U})^2} \partial_{\mu} U \partial^{\mu} \bar{U} \, .
\ee
The complex fields can be written in terms of axions (real parts) and saxions (imaginary parts) as $\tau=\tau_{R}+\mathrm{i} \tau_{I}\,, U=U_{R}+\mathrm{i} U_{I}\,$. From \eqref{kin}, the kinetic terms of these real fields are then
\begin{equation}
\begin{split}
\frac{1}{2} g_{ij} \partial_{\mu} \varphi^i \partial^{\mu} \varphi^j
&= \frac{1}{4 \tau_{I}^2} (\del \tau_{I})^2 +\frac{1}{4 \tau_{I}^2}(\del \tau_{R})^2 + \frac{3}{4 U_{I}^2} (\del U_{I})^2  + \frac{3}{4 U_{I}^2} (\del U_{R})^2  \label{kineticFtheory} \\
&= \frac{1}{2} (\del \hat{\tau}_{I})^2 + \frac{1}{4 e^{2\sqrt{2} \hat{\tau}_{I}}} (\del \hat{\tau}_{R})^2 +\frac{1}{2}  (\del \hat{U}_{I})^2+ \frac{3}{4 e^{\frac{2\sqrt{2}}{\sqrt{3}}\hat{U}_{I}}}  (\del \hat{U}_{R})^2 \,,
\end{split}
\end{equation}
where $\hat{\tau}_{I} = \frac{1}{\sqrt{2}}\log{\tau_{I}}$, $\hat{U}_{I}=\sqrt{\frac{3}{2}} \log{U_{I}}$, and $(\del \tau_{I})^2 \equiv \partial_{\mu} \tau_{I} \partial^{\mu} \tau_{I}$, etc.

Following \cite{Calderon-Infante:2022nxb}, a general 4d scalar potential, in the limits considered, is as given below
\begin{align}\label{Scalarpotential}
V =&\, \frac{1}{\text{vol}} \left( A_{1} e^{- \sqrt{6} \hat{U}_{I} - \sqrt{2}\hat{\tau}_{I}}+A_{2} e^{- \sqrt{\frac{2}{3}} \hat{U}_{I} - \sqrt{2}\hat{\tau}_{I}}+A_{3} e^{ \sqrt{\frac{2}{3}} \hat{U}_{I} - \sqrt{2}\hat{\tau}_{I}}+A_{4} e^{\sqrt{6} \hat{U}_{I} - \sqrt{2}\hat{\tau}_{I}} \right. \\
&\left. +A_{5} e^{-\sqrt{6} \hat{U}_{I} + \sqrt{2}\hat{\tau}_{I}} + A_{6} e^{- \sqrt{\frac{2}{3}} \hat{U}_{I} + \sqrt{2}\hat{\tau}_{I}} +A_{7} e^{ \sqrt{\frac{2}{3}} \hat{U}_{I} + \sqrt{2}\hat{\tau}_{I}}+A_{8} e^{ \sqrt{6} \hat{U}_{I} + \sqrt{2}\hat{\tau}_{I}} +A_{9}-A_{\text{loc}} \right)\,, \nn
\end{align}
where the $A_{i}$ are polynomials of the axions $\tau_R, U_R$ with coefficients given by flux numbers; we detail those in Appendix \ref{ap:CY}. The term $A_{\text{loc}}$ is related to the Euler number of the $CY_{4}$, in such a way as to cancel the tadpole. In \eqref{Scalarpotential}, we wrote for completeness an overall factor proportional to the volume that contains the dependence on the K\"ahler moduli; as mentioned above, we discard this factor in the following.\\

\begin{figure}[H]
\centering
\includegraphics[width=0.6\textwidth]{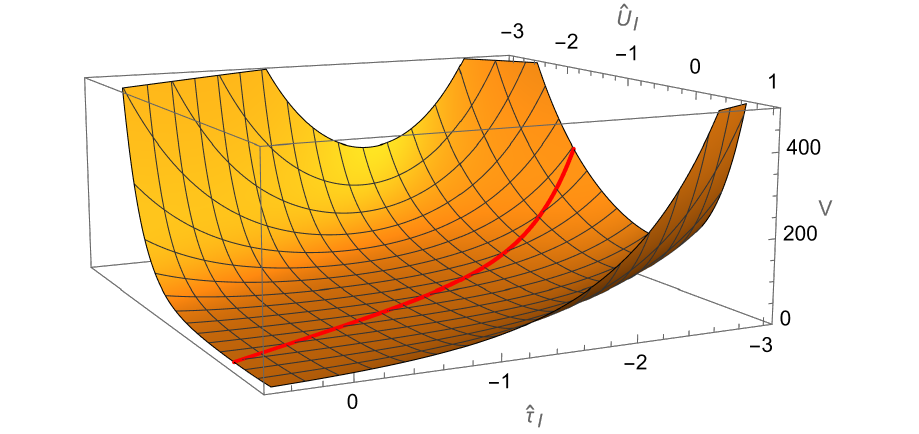}
\caption{$V(\hat{\tau}_{I}\,,\hat{U}_{I})$ for $f_1=f^0=h_0=1$, and the trajectory that goes asymptotically flat in red.}\label{fig:Ftheory}
\end{figure}

We now restrict ourselves to the following choice of flux numbers
\be \label{fluxes}
h^{0}=h^{1}=h_{1} = f^{1} = f_{0} = 0\,,
\ee
which simplifies the $A_i$ detailed in Appendix \ref{ap:CY}. Then the axions only enter the potential as even powers, so they can get stabilised at $\tau_{R} = U_{R} = 0$. The resulting two-field scalar potential reduces to
\be \label{pot}
V(\hat{\tau}_{I}, \hat{U}_{I}) =  (f_{1})^2 \, e^{- \sqrt{\frac{2}{3}} \hat{U}_{I} - \sqrt{2}\hat{\tau}_{I}}+ (f^{0})^{2} \, e^{\sqrt{6} \hat{U}_{I} - \sqrt{2}\hat{\tau}_{I}}+(h_{0})^{2} \, e^{-\sqrt{6} \hat{U}_{I} + \sqrt{2} \hat{\tau}_{I}}\,,
\ee
where as in \cite{Calderon-Infante:2022nxb}, we chose $A_{9}$ such that it cancels $A_{\text{loc}}$. It appears interesting to consider the curve $\hat{U}_{I}=\frac{1}{\sqrt{3}} \hat{\tau}_{I}$, with the asymptotic $\hat{\tau}_{I} \xrightarrow{} \infty$, that corresponds to weak string coupling and large complex structure, i.e.~a limit where the above is trustable. Along this trajectory and in this limit, the potential goes to a positive constant. This is depicted in Figure \ref{fig:Ftheory}.

To make this more manifest, let us define
\be \label{transverse}
\hat{\lambda} = \frac{1}{2}\hat{U}_{I} + \frac{\sqrt{3}}{2} \hat{\tau}_{I}\,,
\quad\quad \hat{\lambda}_{\perp} = - \frac{\sqrt{3}}{2} \hat{U}_{I} + \frac{1}{2} \hat{\tau}_{I}.
\ee
It is easy to check that these fields are canonically normalised, that is, they satisfy
$ (\partial \hat{\tau}_I)^{2} + (\partial \hat{U}_{I})^{2} = (\partial \hat{\lambda})^{2} + (\partial \hat{\lambda}_{\perp})^{2}$. The above scalar potential can then be written in terms of these new fields as follows
 \be \label{interpot}
V = (f_{1})^{2}\, e^{-2\sqrt{\frac{2}{3}}\hat{\lambda}} +(f^{0})^{2}\, e^{-2\sqrt{2} \hat{\lambda}_{\perp}}+(h_{0})^{2}\, e^{2\sqrt{2}\hat{\lambda}_{\perp}} \ .
\ee
The trajectory and limit of interest are now phrased as $\hat{\lambda}_{\bot} = 0 \ ,\ \hat{\lambda} \rightarrow + \infty $, and the scalar potential in this limit goes to
\be \label{finalpot}
V \rightarrow (f^{0})^{2}+(h_{0})^{2}\,.
\ee
It is easy to see that
\beq
\hat{\lambda}_{\perp} = 0 \ ,\ \hat{\lambda} \rightarrow + \infty : \quad \del_{\hat{\lambda}} V \rightarrow 0 \ ,\ V \rightarrow {\rm constant} > 0 \ ,
\eeq
i.e.~the mechanism we \eqref{genmech} is realised here with a slightly more general potential than \eqref{Vgen}, giving an asymptotically vanishing single field slope ratio.

The generalised potential offers an interesting feature: we can also stabilise $\hat{\lambda}_{\perp}$. Indeed, solving $\partial_{\hat{\lambda}_{\perp}}V=0$ gives $\hat{\lambda}_{\perp}=\frac{1}{4\sqrt{2}}\log{\frac{ (f^{0})^2}{ (h_{0})^2}}$, and this is a minimum, similar to the ones studied in \cite{Saltman:2004sn}. For $|f^{0}| =|h_{0}|$, we get the above value for $\hat{\lambda}_{\perp}$, otherwise we can simply redefine $\hat{\lambda}_{\perp}$ by a constant shift. The resulting trajectory then gives asymptotically $\nabla V=0$, i.e.~referring to quantities in \cite{Calderon-Infante:2022nxb}
\beq
\gamma=\epsilon_{V}=0 \ ,
\eeq
recalling that the axions are stabilised and ignoring the K\"ahler moduli. This trajectory was possibly missed in \cite{Calderon-Infante:2022nxb}, since the focus there was on rolling {\sl on-shell} trajectories for which the potential would asymptotically vanish.

\subsubsection{\texorpdfstring{Type IIB and CY\textsubscript{3}}{}}\label{sec:CY3}

We turn to 4d theories coming from compactifications of 10d type IIB supergravity on Calabi-Yau threefolds. More specifically, we study the STU model that arises from the compactification on $T^6/\mathbb{Z}_{2} \times \mathbb{Z}_{2}$, as in \cite{Blaback:2015zra}. The resulting theory is closely related to the previous F-theory example, and as before, we discard the K\"ahler moduli, thus considering {\sl off-shell} trajectories. The K\"ahler potential for the axio-dilaton and the single complex structure modulus considered here is given by
\be
    K = - \log[-\mathrm{i}(\tau - \bar{\tau})] - 3 \log[\mathrm{i}(U-\bar{U})]\,,
\ee
As before, the complex fields can be written in terms of axions (real parts) and saxions (imaginary parts) as $\tau=\tau_{R}+\mathrm{i} \tau_{I}\,, U=U_{R}+\mathrm{i} U_{I}$, and the kinetic terms are identical to the F-theory case \eqref{kineticFtheory}.

The 4d (F-term) scalar potential is given by
\be \label{Ftermpot}
V = e^{K}\left(K^{I \bar{J}} D_{I} W \overline{D_{J}W}-3 |W|^{2}\right)\,,
\ee
where $K_{I\bar{J}}=\partial_{I}\partial_{\bar{J}}K\,$, and where the superpotential is given as follows for this simple STU model \cite{Blaback:2015zra}
\be \label{GKWIIB}
     W =(f^{0} - \tau h^{0}) U^3- 3(f^{1}-\tau h^{1}) U^2+ 3 (f_{1}-\tau h_{1}) U + (f_{0}- \tau h_{0})\,,
\ee
adapted to our conventions. The scalar potential is straightforward to obtain, even though complicated, and more general than the previous one \eqref{Scalarpotential}. The structure remains similar, in particular the axions appear within polynomials with flux numbers as coefficients. The latter are in addition constrained to obey the tadpole cancellation condition \cite{Lust:2006zh}
 \be \label{tadpole}
\int F_{3} \wedge H_{3} = \frac{1}{2} N_{O3} = 16\,.
 \ee

We now make the following choices for the fluxes
\be
h^{0}=h^{1}=h_{1}=f_{1}=0\,\,,
\quad\quad
f^{0}=\frac{16}{h_{0}}\,,
\ee
which simplify the scalar potential. From the latter, one can stabilise the axions at the following values
\be
\tau_{R} =\frac{f_{0}}{h_{0}}+\frac{(f^{1})^{3} h_{0}}{3456}\,,
\quad\quad
U_{R} = \frac{f^{1} h_{0}}{48}\,.
\ee
The resulting two-field scalar potential is given as follows
\begin{equation}
\begin{split}
V(\hat{\tau}_{I}, \hat{U}_{I})=&\, 14 - \frac{(f^{1})^2}{8} e^{-\sqrt{2}\hat{\tau}_{I}+\sqrt{\frac{2}{3}}\hat{U}_{I}} + \frac{16}{(h_{0})^2} e^{-\sqrt{2}\hat{\tau}_{I}+\sqrt{6} \hat{U}_{I}} +\frac{(h_{0})^{2}}{16} e^{\sqrt{2}\hat{\tau}_{I}-\sqrt{6} \hat{U}_{I}} \\
&+\frac{(f^{1} h_{0})^2 }{128} e^{-2 \sqrt{\frac{2}{3}} \hat{U}_{I}} - \frac{5(f^{1})^{4} (h_{0})^2 }{110592} e^{-\sqrt{2}\hat{\tau}_{I}-\sqrt{\frac{2}{3}} \hat{U}_{I}}\,.
\end{split}
\end{equation}
In view of identifying a trajectory of interest, we now define
 \be
\hat{\lambda} = \frac{1}{2}\hat{U}_{I} + \frac{\sqrt{3}}{2} \hat{\tau}_{I}
\,,\quad\quad \hat{\lambda}_{\perp} = - \frac{\sqrt{3}}{2} \hat{U}_{I} + \frac{1}{2} \hat{\tau}_{I} \ ,
\ee
a new pair of canonically normalised fields, namely $ (\partial \hat{\tau}_I)^{2} + (\partial \hat{U}_{I})^{2} = (\partial \hat{\lambda})^{2} + (\partial \hat{\lambda}_{\perp})^{2}$. Note that as above, the limit $\hat{\lambda} \rightarrow + \infty$ is one where the 4d theory can be trusted. The scalar potential gets rewritten as
\begin{equation}
\begin{split}
V(\hat{\lambda},\hat{\lambda}_{\perp})=&\,
14 + \frac{16}{(h_{0})^2} e^{-2\sqrt{2}\hat{\lambda}_{\perp}} +\frac{(h_{0})^{2}}{16} e^{2\sqrt{2}\hat{\lambda}_{\perp}}\\
& + \frac{(f^{1})^2}{8} e^{-\sqrt{\frac{2}{3}}\hat{\lambda}} \left( \frac{(h_{0})^2 }{16}e^{\sqrt{2}\hat{\lambda}_{\perp}} -e^{-\sqrt{2}\hat{\lambda}_{\perp}} \right) - \frac{5(f^{1})^{4} (h_{0})^2 }{110592} e^{-2\sqrt{\frac{2}{3}}\hat{\lambda}} \,.
\end{split}
\end{equation}
It is now easy to see that the following trajectory and asymptotic gives the desired result
\beq
\hat{\lambda}_{\perp} = 0 \ ,\ \hat{\lambda} \rightarrow + \infty : \quad \del_{\hat{\lambda}} V \rightarrow 0 \ ,\ V \rightarrow {\rm constant} > 0 \ ,
\eeq
realising once again the mechanism \eqref{genmech} with a potential more general than \eqref{Vgen}.

\begin{figure}[H]
\centering
\includegraphics[width=0.6\textwidth]{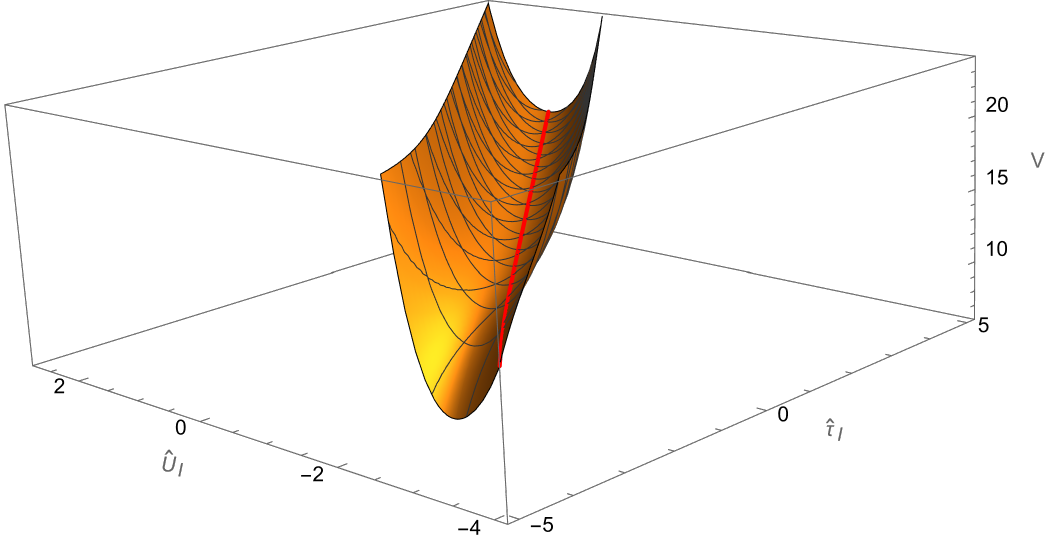}
\caption{$V(\hat{\tau}_{I}\,,\hat{U}_{I})$ with $f^1=h_0=1$, and the trajectory that goes asymptotically flat in red, for which we make the stabilising shift $\hat{\lambda}_{\perp} \rightarrow \hat{\lambda}_{\perp} +\frac{1}{2\sqrt{2}}\log{\frac{16}{h_{0}^2}}$.}\label{fig:IIB}
\end{figure}

In the limit $\hat{\lambda} \xrightarrow{} \infty$, we can stabilise $\hat{\lambda}_{\perp}$ by solving $\partial_{\hat{\lambda}_{\perp}}V=0$: we get the value $\hat{\lambda}_{\perp}=\frac{1}{2\sqrt{2}}\log{\frac{16}{h_{0}^2}}$. Note that the flux $h_{0}$ is constrained by the tadpole, and can only take a discrete set of values. $\hat{\lambda}_{\perp}$ can be redefined by a constant shift to define the trajectory at its stabilised value, instead of $\hat{\lambda}_{\perp}=0$. We would then get once again $\epsilon_V \rightarrow 0$ in the limit considered, ignoring the K\"ahler moduli. We illustrate our results in Figure \ref{fig:IIB}.

\subsection{Landau-Ginzburg model}\label{sec:LG}

We turn to the 4d effective theory obtained from the orientifold of the Landau-Ginzburg orbifold $1^9$ model. We review the latter in Appendix \ref{ap:LG}. This string model is dubbed ``non-geometric'' since it does not provide a standard 10d target space geometry, with a 6d compact manifold; it nevertheless gives rise to a 4d effective theory, formally similar to that obtained through Calabi-Yau compactifications. It admits no K\"ahler moduli: since those are related to volumes of cycles in the compact manifold, having no K\"ahler modulus highlights the ``non-geometric'' nature of such model.

The 4d effective theory admits ${\cal N}=1$ supersymmetry: it is then formulated as above thanks to a K\"ahler potential and a superpotential. As explained in Appendix \ref{ap:LG}, we consider the large complex structure limit of the untwisted sector, such that we only get a dependance on the three untwisted complex structure moduli, together with the axio-dilaton \cite{Becker:2007dn}. The K\"ahler potential and superpotential are then given by
\bea \label{Kaehlerpotential}
K &=& -4 \log\left[-\mathrm{i} (\tau-\bar{\tau})\right]- \log \left[\mathrm{i} (U_{1}-\bar{U}_{1})(U_{2}-\bar{U}_{2})(U_{3}-\bar{U}_{3})\right]\,,\\[2mm]
W &=& (f^{0} - \tau h^{0}) U_{1} U_{2} U_{3} - (f^{1}-\tau h^{1}) U_{2} U_{3}-(f^{2}-\tau h^{2}) U_{1} U_{3}-(f^{3}-\tau h^{3}) U_{1} U_{2} \nonumber\\
&&+ (f_{1}-\tau h_{1}) U_{1}+(f_{2}-\tau h_{2}) U_{2}+(f_{3}-\tau h_{3}) U_{3} + (f_{0}- \tau h_{0})\,. \label{GKW}
\eea
The complex fields are composed of axions (real parts) and saxions (imaginary parts): $\tau = \tau_{R}+\mathrm{i} \tau_{I}$, $U_{1}=U_{1R}+\mathrm{i} U_{1I}$, $U_{2}=U_{2R}+\mathrm{i} U_{2I}$, $U_{3}=U_{3R}+\mathrm{i} U_{3I}$. As above, the kinetic terms for the saxions and axions are then
\begin{align}\label{Kinetic}
   \frac{1}{2}g_{ij}\partial_{\mu}\varphi^{i} \partial^{\mu} \varphi^{j}
   &= \frac{1}{\tau_{I}^{2}} (\del \tau_{I})^2 + \frac{1}{\tau_{I}^{2}} (\del \tau_{R})^2 +\frac{1}{4 U_{1 I}^{2}} (\del U_{1 I})^2+\frac{1}{4 U_{2 I}^{2}} (\del U_{2 I})^2 + \frac{1}{4 U_{3 I}^{2}} (\del U_{3 I})^2 \nn\\
   &\,\,\,\,\,\, +\frac{1}{4 U_{1 I}^{2}} (\del U_{1 R})^2 +\frac{1}{4 U_{2 I}^{2}} (\del U_{2 R})^2 +\frac{1}{4 U_{3 I}^{2}} (\del U_{3 R})^2 \nn \\
   &=\frac{1}{2} (\del \hat{\tau}_{I})^2 +\frac{1}{2} (\del \hat{U}_{1I})^2 +\frac{1}{2} (\del \hat{U}_{2I})^2 +\frac{1}{2} (\del \hat{U}_{3I})^2 \\
   &\,\,\,\,\,\,+\frac{1}{e^{\sqrt{2} \hat{\tau}_{I}}} (\del \tau_{R})^2 +\frac{1}{4 e^{2 \sqrt{2} \hat{U}_{1 I}}} (\del U_{1 R})^2 +\frac{1}{4 e^{2 \sqrt{2} \hat{U}_{2 I}}} (\del U_{2 R})^2+\frac{1}{4 e^{2 \sqrt{2} \hat{U}_{3 I}}} (\del U_{3 R})^2\,,\nn
\end{align}
where $\hat{\tau}_{I}=\sqrt{2} \log\tau_{I}$\,, $\hat{U}_{1I}=\frac{1}{\sqrt{2}}\log U_{1I}\,,\hat{U}_{2I}=\frac{1}{\sqrt{2}}\log U_{2I}\,, \hat{U}_{3 I} = \frac{1}{\sqrt{2}}\log U_{3I}$\,. The scalar potential is also straightforward to derive and we give it in \eqref{LGpot}, following \cite[(3.12)]{Cremonini:2023suw}. The axions appear again in polynomial functions with flux numbers as coefficients. The latter are subject to the tadpole cancellation condition, given by
 \be
\int F_{3} \wedge H_{3} = \frac{1}{2} N_{O3} = 12 \ .
 \ee

\begin{figure}[H]
\centering
\includegraphics[width=0.6\textwidth]{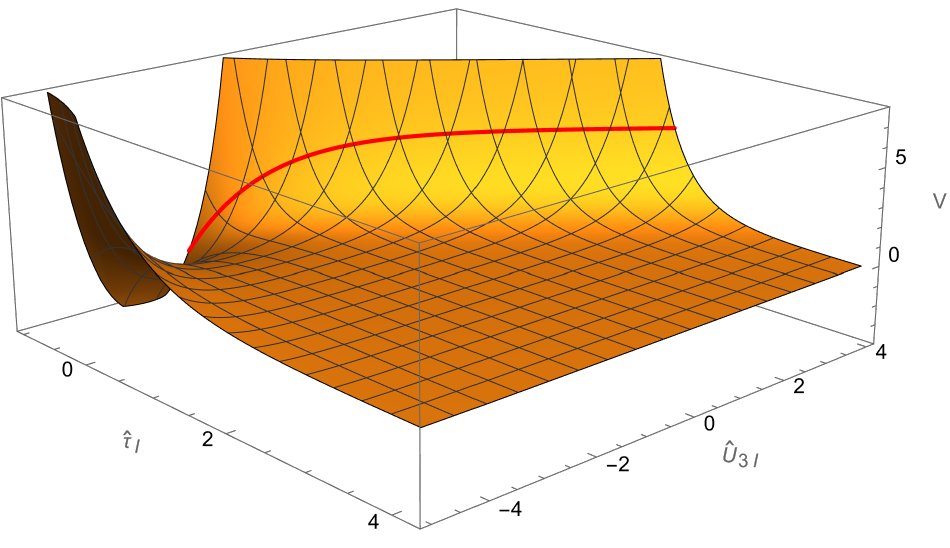}
\caption{$V(\hat{\tau}_{I},\, \frac{1}{2}\hat{U}_{3I},\, \frac{1}{2}\hat{U}_{3I},\, \hat{U}_{3I})$, and the trajectory of interest in red.}\label{fig:LG}
\end{figure}

We now make the following (allowed) choice of fluxes entering the potential \eqref{LGpot}
\be
h_0=h_2=h^0=h^1=h^2=h^3=f_0=f^0=f_1=f_2=f_3=f^2=f^3=0\ ,\ f^1=\frac{12}{h_1} \ ,
\ee
for which the tadpole cancellation condition is satisfied. Then we stabilise the axions at the following values
\be \label{axions}
\tau_{R} = U_{1R} = U_{2R}=U_{3R}=0\,.
\ee
The scalar potential reduces to
\begin{equation}
\begin{split}
   V =&\, \frac{9}{2(h_1)^2} e^{\sqrt{2}\,(-2\hat{\tau}_{I}-\hat{U}_{1 I}+\hat{U}_{2 I}+\hat{U}_{3 I})} - \frac{3}{4} e^{-\frac{3}{\sqrt{2}}\hat{\tau}_{I}}  + \frac{(h_1)^2}{128} e^{\sqrt{2}\,(-\hat{\tau}_{I}+\hat{U}_{1 I}-\hat{U}_{2 I}-\hat{U}_{3 I})}  \\
   & + \frac{h_{3}^{2}}{128} e^{\sqrt{2}\,(-\hat{\tau}_{I}-\hat{U}_{1 I}-\hat{U}_{2 I}+\hat{U}_{3 I})}   - \frac{3 h_{1} h_{3}}{64} e^{\sqrt{2}\,(-\hat{\tau}_{I}-\hat{U}_{2 I})}\,.
\end{split}
\end{equation}
With respect to previous examples, we now face a four-field potential. To make apparent the direction of interest, we perform a change of basis $\vec{\hat{\lambda}}= M \vec{\hat{\varphi}}$ where $\vec{\hat{\varphi}}=\{\hat{\tau}_{I},\hat{U}_{1 I},\hat{U}_{2 I},\hat{U}_{3 I}\}$\,,\,\,$\vec{\hat{\lambda}}=\{\hat{\lambda}_{1},\hat{\lambda}_{2}, \hat{\lambda}_{3},\hat{\lambda}_{4}\}$ and the transformation matrix is given by
\be
M=\left(
\begin{array}{cccc}
 \frac{2}{\sqrt{7}} & \frac{1}{\sqrt{7}} &
   -\frac{1}{\sqrt{7}} & -\frac{1}{\sqrt{7}} \\
 \frac{1}{\sqrt{7}} & \frac{1}{\sqrt{7}} &
   \frac{1}{\sqrt{7}} & \frac{2}{\sqrt{7}} \\
 -\sqrt{\frac{2}{7}} & \frac{3}{\sqrt{14}} &
   -\frac{1}{\sqrt{14}} & 0 \\
 0 & \frac{1}{\sqrt{14}} & \frac{3}{\sqrt{14}} &
   -\sqrt{\frac{2}{7}} \\
\end{array}
\right)\,.
\ee
It is easy to see that the above matrix is orthonormal, so the new fields are also canonically normalised. In terms of the old fields, the trajectory we are interested in is $\hat{\tau}_{I}=\hat{U}_{1 I}=\hat{U}_{2 I}=\frac{1}{2}\hat{U}_{3 I}=\sigma$ where $\sigma \xrightarrow{} \infty$. In the new basis this would translate into $\hat{\lambda}_{1}=\hat{\lambda}_{3}=\hat{\lambda}_{4}=0\,,\hat{\lambda}_{2}=\sqrt{7}\, \sigma$ where $\sigma \xrightarrow{} \infty$: this defines the trajectory and asymptotic of interest here. In the new basis, the potential is given by
\begin{equation}
\begin{split}
V =&\,\, \frac{9}{2(h_1)^2} e^{-\sqrt{14}\hat{\lambda}_{1}} - \frac{3}{4} e^{-\frac{3}{\sqrt{14}}(2\hat{\lambda}_{1}+\hat{\lambda}_{2}-\sqrt{2} \hat{\lambda}_{3})}  + \frac{(h_1)^2}{128} e^{\frac{\sqrt{2}\hat{\lambda}_{1}-3 \sqrt{2}\hat{\lambda}_{2}+6 \hat{\lambda}_{3}}{\sqrt{7}}}  \\
& + \frac{h_{3}^{2}}{128} e^{\frac{-3\sqrt{2}\hat{\lambda}_{1}- \sqrt{2}\hat{\lambda}_{2}-6 \hat{\lambda}_{4}}{\sqrt{7}}}   - \frac{3 h_{1} h_{3}}{64} e^{\frac{-\sqrt{2} \hat{\lambda}_{1}-2 \sqrt{2} \hat{\lambda}_{2} +3 (\hat{\lambda}_{3}-\hat{\lambda}_{4})}{ \sqrt{7}}}\,.
\end{split}
\end{equation}
It is then easy to see that the mechanism \eqref{genmech} is realised as follows
\beq
\hat{\lambda}_{1}=\hat{\lambda}_{3}=\hat{\lambda}_{4}=0\,,
\quad\hat{\lambda}_{2}\rightarrow + \infty : \quad \del_{\hat{\lambda}_2} V \rightarrow 0 \,,
\quad V \rightarrow {\rm constant} > 0 \,.
\eeq
Note we also get $\del_{\hat{\lambda}_3} V,\,  \del_{\hat{\lambda}_4} V \rightarrow 0$ in the limit considered. The above is a four-field example of an asymptotically vanishing single field slope ratio. We illustrate these results in Figure \ref{fig:LG}.

\subsection{DGKT compactification}\label{sec:DGKT}

We focus here on a well-known example of a 4d string effective theory, obtained from the compactification of 10d massive type IIA supergravity on the toroidal orbifold $T^6/\mathbb{Z}_3\times\mathbb{Z}_3$. The setting admits the so-called DGKT anti-de Sitter solution \cite{DeWolfe:2005uu} (see also \cite{Lust:2004ig, Camara:2005dc, Acharya:2006ne}), which is recovered as a negative minimum of the 4d scalar potential. We briefly review in Appendix \ref{ap:DGKT} the derivation of the 4d theory to be used here, following conventions of \cite{Andriot:2023fss}, as well as the values obtained at this negative minimum.

The 4d theory to be considered in this subsection, of the form \eqref{action}, has four scalar fields: two saxions $g,r$ related to the dilaton and the (isotropic) metric deformation, and two axions $\tilde{b},\tilde{\xi}$ coming from the 10d fields $B_2, C_3$. As detailed in Appendix \ref{ap:DGKT}, we focus here for simplicity on this isotropic set of fields; the non-isotropic case, giving rise to eight fields, is given in the appendix. We therefore have the following kinetic terms
\begin{align}\label{kiniso}
    S_{kin}
    &=-\frac{M_p^2}{2}\int\text{d}^4x\sqrt{-g_{4}}\left(
    \frac{2}{g^2}(\partial g)^2
    +\frac{6}{r^2}(\partial r)^2
    +\frac{3}{r^4}(\partial  \tilde{b})^2
    +2g^2(\partial \tilde{\xi})^2\right)\,,
\end{align}
while the scalar potential is proportional to
\begin{equation}\label{potpot}
\begin{split}
    V\sim
    \frac{1}{2}\Biggl(&
    \frac{1}{2}\frac{g^2}{r^6}
    -2\sqrt{2}g^3
    +3\frac{g^4}{r^2}
    +g^4r^6
    +\frac{g^4}{r^6}\Big(6\tilde{b}^2r^8
    -4\tilde{b}^2r^4\sum_{i=1}^3s_i
    +2(\tilde{\xi}+\tilde{b}\sum_{i=1}^3s_i)^2\Big)\\
    &
    +\frac{g^4}{r^6}\Big(12\tilde{b}^4r^4
    -8\tilde{b}^3(\tilde{\xi}+\tilde{b}\sum_{i=1}^3s_i)
    \Big)
    +8\frac{g^4}{r^6}\tilde{b}^6\Biggl) \,,
\end{split}
\end{equation}
where we have ignored an overall factor $(2\pi\sqrt{\alpha^{\prime}})^6 M_p^2 \ \frac{p^4\vert m_0\vert^{5/2}}{E^{3/2}}$. Note that, as detailed in Appendix \ref{ap:DGKT}, we computed here the complete expression of the potential, including quartic and sextic axionic terms, with the motivation of going away from the vacuum.\\

We set for now the axions to their vacuum expectation value, $\tilde{b}=\tilde{\xi}=0$, and focus on the resulting two-field theory. It is easy to see that the classical limit, in which the theory can be trusted, corresponds to $r \rightarrow \infty,\ g\rightarrow 0$ (we recall that $g \propto e^{\phi}/r^3$). In this limit, the only potential term that can asymptote to a non-zero constant is $g^4 r^6$, indicating the trajectory of interest to be $g \sim r^{-\frac{3}{2}} $. Introducing canonical fields via $r= e^{\frac{\hat{r}}{\sqrt{6}M_p}}$, $g= e^{\frac{\hat{g}}{\sqrt{2}M_p}}$, the limit of interest becomes $\hat{r}\rightarrow \infty, \hat{g}\rightarrow -\infty$; in the following we set $M_p=1$ for simplicity. The potential dominant term becomes proportional to $e^{2\sqrt{2}\hat{g}+\sqrt{6}\hat{r}}$, and the curve of interest is $\hat{g}=-\frac{\sqrt{3}}{2}\,\hat{r}$ (up to a possible constant shift).

\begin{figure}[H]
\centering
\includegraphics[width=0.6\textwidth]{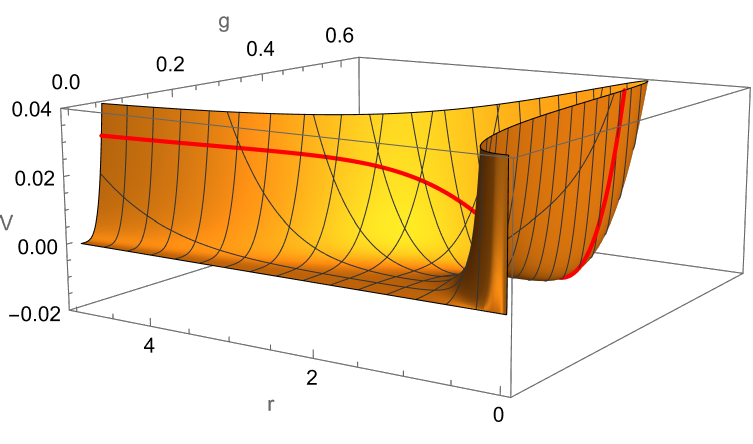}
\caption{$V(g,r)$ for the DGKT compactification, with the anti-de Sitter minimum visible, and the trajectory of interest that goes asymptotically flat, in red. We adjusted the latter towards $g/{\langle g \rangle} = (r/{\langle r \rangle})^{-3/2}$, in order for it to pass by the minimum; the fields vev are given in Appendix \ref{ap:DGKT}. This amounts to a constant shift in the definition of canonical fields, which does not change the result.}\label{fig:DGKT}
\end{figure}

So we introduce the following new pair of fields
\begin{align}
    \hat{\lambda}
    =
    -\frac{\sqrt{3}}{2\alpha}\hat{g}
    +\frac{1}{\alpha}\hat{r}
    \,,&\quad\quad
    \hat{\lambda}_{\bot}
    =
    -\frac{1}{\alpha}\hat{g}
    -\frac{\sqrt{3}}{2\alpha}\hat{r} \,,\\
   \leftrightarrow
   \hspace{0.5cm}\hat{g} =  -\frac{\alpha}{7} \left( 2 \sqrt{3} \hat{\lambda} + 4 \hat{\lambda}_{\bot} \right)   \,,&\quad\quad \hat{r} =  \frac{\alpha}{7} \left( 4 \hat{\lambda} - 2 \sqrt{3} \hat{\lambda}_{\bot} \right) \,,
\end{align}
where $\alpha=\sqrt{7}/2$, such that ones verifies $(\partial_{\mu}\hat{\lambda})^2+(\partial_{\mu}\hat{\lambda}_{\bot})^2 = (\partial_{\mu}\hat{g})^2+(\partial_{\mu}\hat{r})^2$. The trajectory of interest is now along $\hat{\lambda}$ with $\hat{\lambda}_{\bot}= 0$, with the (classical) limit $\hat{\lambda}\rightarrow +\infty$. The complete two-field (saxionic) potential gets rewritten, from \eqref{potpot}, into
\begin{equation}
\begin{split}
    V(\hat{\lambda} ,\hat{\lambda}_{\bot})\sim &\
    \frac{1}{2} e^{-\sqrt{14}\, \hat{\lambda}_{\bot}} \\
    & + \frac{1}{4} e^{-\sqrt{\frac{2}{7}}\left( 3 \sqrt{3} \hat{\lambda} - \hat{\lambda}_{\bot} \right)} - \sqrt{2} e^{-\frac{1}{\sqrt{14}} \left( 3 \sqrt{3} \hat{\lambda} + 6 \hat{\lambda}_{\bot} \right)} + \frac{3}{2} e^{-\frac{1}{\sqrt{14}} \left(\frac{16}{\sqrt{3}} \hat{\lambda} + 6 \hat{\lambda}_{\bot}  \right)} \,.
\end{split}
\end{equation}
It is straightforward to verify that this realises the general mechanism \eqref{genmech}
\beq
\hat{\lambda}_{\bot} = 0 \ ,\ \hat{\lambda} \rightarrow + \infty : \quad \del_{\hat{\lambda}} V \rightarrow 0 \ ,\ V \rightarrow {\rm constant} > 0 \ ,
\eeq
giving an asymptotically vanishing single field slope ratio. We illustrate our results in Figure \ref{fig:DGKT}.

Let us emphasize that our analysis puts forward the positive part of the DGKT scalar potential, which is not often considered. In particular, Figure \ref{fig:DGKT} exhibits a continuous trajectory that connects the well-known negative minimum of the potential to its positive part. This is done along a classical direction, on a wall (or steep slope) due to the $g^4 r^6$ term.\\

\begin{figure}[H]
\centering
\includegraphics[width=0.6\textwidth]{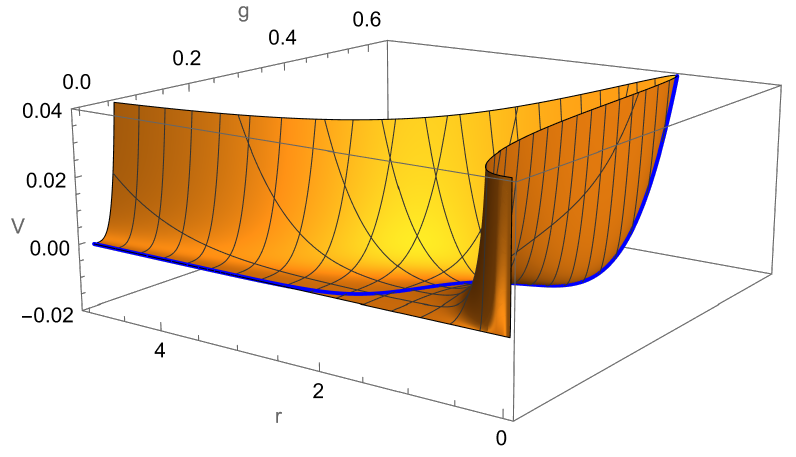}
\caption{$V(g,r)$ for the DGKT compactification as in Figure \ref{fig:DGKT}, and the trajectory discussed in the main text, in blue, for which $V\rightarrow 0$. We adjusted the latter towards $g/{\langle g \rangle} = (r/{\langle r \rangle})^{-6}$, in order for it to pass by the minimum.}\label{fig:DGKTV0}
\end{figure}

Another trajectory is better known in the DGKT 4d theory: as mentioned already in the original work \cite{DeWolfe:2005uu}, considering $g \sim r^{-6}$ can connect the vacuum to a classical asymptotic $g\rightarrow 0, r\rightarrow \infty$ where $V\rightarrow 0$. Let us briefly discuss this trajectory here, building on \cite[Sec. 5.2.1]{Andriot:2022brg} where it was partially treated. We depict this trajectory in Figure \ref{fig:DGKTV0}. The curve $g \sim r^{-6}$ corresponds, in terms of canonical fields, to the trajectory $\hat{g} = -2\sqrt{3} \hat{r}$, ignoring again for simplicity a constant shift. We then define the following canonical fields
\begin{align}
\hat{\lambda} = - \frac{2\sqrt{3}}{\sqrt{13}} \, \hat{g} + \frac{1}{\sqrt{13}} \, \hat{r} \ , & \quad \hat{\lambda}_{\bot} = \frac{1}{\sqrt{13}} \,\hat{g} + \frac{2\sqrt{3}}{\sqrt{13}} \, \hat{r} \\
\hspace{0.5cm}\leftrightarrow \hat{g} = - \frac{2\sqrt{3}}{\sqrt{13}} \hat{\lambda} + \frac{1}{\sqrt{13}} \hat{\lambda}_{\bot} \ , & \quad \hat{r} = \frac{1}{\sqrt{13}} \hat{\lambda} + \frac{2\sqrt{3}}{\sqrt{13}} \hat{\lambda}_{\bot} \ ,
\end{align}
such that $(\partial \hat{\lambda})^2+(\partial \hat{\lambda}_{\bot})^2=(\partial \hat{g})^2+(\partial \hat{r})^2$. We rewrite the saxionic potential from \eqref{potpot} as follows\footnote{The difference in the multiplicative constants between \eqref{VusualdirDGKT} and the potential in \cite[Sec. 5.2.1]{Andriot:2022brg} is due to the freedom in adding constants to the canonical field definitions, while still maintaining the kinetic terms; the exponential rates remain identical, which is what matters here.}
\beq
V \sim e^{ -3\sqrt{\frac{6}{13}}\, \hat{\lambda}} \left( \frac{1}{4} e^{-5\sqrt{\frac{2}{13}} \, \hat{\lambda}_{\bot}  } - \sqrt{2}\, e^{ \frac{3}{2} \sqrt{\frac{2}{13}}  \, \hat{\lambda}_{\bot}}  + \frac{1}{2} e^{ 8\sqrt{\frac{2}{13}} \, \hat{\lambda}_{\bot} } \right)+ \frac{3}{2} e^{ -\sqrt{\frac{26}{3}} \hat{\lambda}   } \ ,\label{VusualdirDGKT}
\eeq
and the dominant term in the limit of interest ($\hat{\lambda}_{\bot}=0 \,,
\ \hat{\lambda} \rightarrow \infty $) is the first one. In these asymptotics, we conclude on the following behaviour
\beq
\hat{\lambda}_{\bot}=0 \,,
\quad \hat{\lambda} \rightarrow \infty \,:
\quad \del_{\hat{\lambda}} V \rightarrow 0 \,,
\quad V \rightarrow 0 \,,
\quad \frac{\del_{\hat{\lambda}} V}{V} \rightarrow -3\sqrt{\frac{6}{13}} \, .
\eeq
The single field slope ratio along this well-known trajectory therefore does not vanish asymptotically (it even obeys the (A)TCC bound \cite{Andriot:2022brg}). This is due to the potential being exponential, and vanishing itself in the limit.

Let us finally add a word on the axionic sector: is it possible to consider other trajectories where $\tilde{b}$ or $\tilde{\xi} \rightarrow \infty$ ? We recall that the background fluxes break the standard axionic shift symmetry, making their field space non-compact, hence the possibility of such asymptotics.

Let us stick to a two-field theory with {\sl off-shell} trajectories. If we set $g,r$ to given values, the axionic potential displays a $\tilde{b}^6$ term and $\tilde{\xi}^2$ term, both of which would diverge in the limit considered. If we want to set only $g$ or $r$ to a given value, it has to be $r$: otherwise, the term $\langle g \rangle^4 r^6$ would diverge in the classical limit. We are left with fixing $r$ and another axion, say $\tilde{\xi}$, while considering the two fields $g, \tilde{b}$. In that case, the dominant term, in the limit $g\rightarrow 0, \, \tilde{b} \rightarrow \infty$, is $g^4 \tilde{b}^6$ which could be brought to a positive constant.

We would then face the problem of the kinetic terms mentioned in Appendix \ref{ap:DGKT}: the axion field kinetic terms cannot be canonically normalized, due to their dependence in the saxions. However, placing ourselves in an {\sl off-shell} situation where we fix $r$ and $\tilde{\xi}$, this problem seems to go away. In terms of seemingly canonically normalized fields, the potential term is then of the form $e^{2\sqrt{2}\hat{g}}\, \tilde{b}^6$, which can be sent to a constant along the curve $e^{2\sqrt{2}\hat{g}} = \tilde{b}^{-6}$. It is however difficult to identify a canonical field basis that would capture that curve, as done for previous examples, and therefore to compute the single field slope ratio. We refrain for now from further investigation with axions.

\section{On-shell potential characterisation}\label{sec:onshell}

In previous sections, we have shown that achieving an {\sl off-shell} characterisation of a stringy scalar potential in a multifield setting is delicate. Indeed, considering trajectories as those depicted in Figure \ref{fig:intro} leads in \eqref{question1} to having a vanishing lower bound, $c=0$, to the (asymptotic) single field slope ratio $\frac{| \del_{\hat{\lambda}} V |}{V}$. Such a bound gives an empty condition, and the SdSC characterisation \eqref{SdSC} thus seems better suited. However, trajectories of Figure \ref{fig:intro} are not physical, as we will detail below. This leads to the question \eqref{question2}, whether by restricting to physical or {\sl on-shell} trajectories, we may circumvent the above conclusion and get a characterisation of the potential in a multifield setting, that differs from the SdSC.\footnote{Let us recall that the SdSC was initially considered valid in asymptotics of field space but also in time, the latter referring to a physical trajectory (as well as to fixed points in the evolution of the physical system) \cite{Rudelius:2021oaz, Rudelius:2021azq}. Also, the TCC was originally derived using the (single) field equation of motion, therefore referring again to a physical trajectory \cite{Bedroya:2019snp}. Finally, the counter-example to the latter found in \cite{Calderon-Infante:2022nxb} was obtained along a physical trajectory, as we will discuss, given one ignores the K\"ahler moduli. Note also a recent (single field) proposal \cite{Nitta:2025mpi} including kinetic energy along the physical trajectory. Therefore, as explained in the Introduction, considering characterisations along {\sl on-shell} trajectory is not a new idea, however those obtained were often promoted to {\sl off-shell} ones.}\\

Let us first discuss the terminology: {\sl on-shell} refers to field values that can be obtained by a physical evolution. The simplest realisation is of course a trajectory in field space which is dictated by the classical equations of motion, given some physical initial conditions; in the following we will stick to the notion of trajectory and refer to equations of motion, even when its ingredients (e.g.~the potential or field space metric) receive corrections beyond a classical regime. However, we know that non-perturbative or quantum phenomena can provide different evolutions, such as tunneling. The notion of field value and trajectory is then not necessarily well-defined. We will set such situations aside here, and assume that the physics evolve in a regime suited to our description, which should happen when focusing on appropriate asymptotic regions.

In addition, we would like to restrict to cosmological backgrounds, namely FLRW metrics, and to time-dependence only, i.e.~$\varphi^i(t)$ and $\dot{\varphi}^i \equiv \del_t \varphi^i$. This leads to the following equations of motion in $d$ dimensions, $d \geq 3$,
\begin{align}
H^2 &= \frac{2}{(d-1)(d-2)} \left( \sum_i \rho_i + \frac{1}{2} g_{ij} \dot{\varphi}^i \dot{\varphi}^j + V \right) \ , \label{Friedmann}\\
0 &=  \ddot{\varphi}^i + \Gamma_{jk}^i \dot{\varphi}^j \dot{\varphi}^k + (d-1) H \dot{\varphi}^i + g^{ij} \del_{\varphi^i} V \ ,\label{fieldeq}
\end{align}
while the other Friedmann equation is obtainable from those two for $H\neq0$. $H$ is the Hubble parameter, and we take $H>0$ for an expanding universe. The $\rho_i$ denote other possible energy densities (e.g.~matter, radiation), coming in addition to \eqref{action}; for simplicity here, we assumed them to be independent of the $\varphi^i$, i.e.~we considered no direct coupling. Finally, $\Gamma_{jk}^i$ is the Christoffel symbol for the field space metric $g_{ij}$. In the following, ``physical'' or ``{\sl on-shell} trajectories'' will be those dictated by these equations.

The field equation \eqref{fieldeq} contains two important terms: the Hubble friction, which decelerates the field, and the force due to the potential slope, which accelerates the field towards the bottom of the potential. In all examples presented in Section \ref{sec:examples}, we had a non-zero potential gradient: indeed, even when the transverse field considered got stabilized asymptotically, as in Figure \ref{fig:Ftheory} and \ref{fig:IIB}, other discarded fields (the K\"ahler moduli) would be rolling. Therefore, let us consider, as e.g.~in Figure \ref{fig:intro}, a canonical transverse field $\hat{\lambda}_{\bot}$ for which $\del_{\hat{\lambda}_{\bot}} V < 0$. Assume initial conditions where the fields follow the would-be ``{\sl off-shell} trajectory'' with $\hat{\lambda}_{\bot}=0$ (e.g.~the red trajectory in Figure \ref{fig:intro}), meaning initial conditions where $\dot{\hat{\lambda}}_{\bot} = 0$. The transverse field equation, given by
\beq
\ddot{\hat{\lambda}}_{\bot} = - (d-1) H \dot{\hat{\lambda}}_{\bot} - \del_{\hat{\lambda}_{\bot}} V \ ,
\eeq
indicates that initially, $\ddot{\hat{\lambda}}_{\bot}>0$. This implies that the field gains some positive speed, so $\hat{\lambda}_{\bot}$ grows. We have just mathematically described the obvious: if the fields try to follow the ``{\sl off-shell} trajectory'' as in Figure \ref{fig:intro}, they soon deviate from it by rolling down the potential along $\hat{\lambda}_{\bot}$. Different initial conditions mean $\dot{\hat{\lambda}}_{\bot} \neq 0$, so they do not change the fact that the ``{\sl off-shell} trajectory'' is immediately left. This shows that the fields, in such a physical evolution, never follow the asymptotically flat directions previously discussed.

At best, the fields are close to these asymptotically flat trajectories during a transient phase, before moving away; as a side comment, note that such a transient phase may still be phenomenologically relevant to the present universe acceleration. This situation shows that statements on a single field slope ratio $\frac{|\del_{\hat{\lambda}} V|}{V}$ cannot be made locally, even for physical trajectories, since the former can locally vanish. A statement on $\frac{|\del_{\hat{\lambda}} V|}{V}$ may at best be possible in the asymptotics of physical trajectories: let us now discuss this possibility.\\

{\sl What can be said about asymptotics of physical trajectories?} One could think that because of friction, the fields slow down and eventually follow the steepest descent trajectory; as we will see, this is only true upon some extra restrictions. Following the steepest descent of a potential amounts to follow a gradient flow (see e.g.~\cite[Sec. 2.2.2]{Andriot:2023isc} and references therein): this means that the speed vector (by definition, tangent to the trajectory) is proportional to the gradient, i.e.~in components
\beq
\dot{\varphi}^i \propto g^{ij} \del_{\varphi^j} V \ ,
\eeq
where the factor can be a function. From the field equation \eqref{fieldeq}, it is easy to see that the slow-roll approximation, which neglects the first two terms against the others, gives the friction term proportional the potential derivative, realising a gradient flow \cite{Calderon-Infante:2022nxb}. The latter might be understood intuitively: at low speeds, one can imagine that the trajectory will get oriented by the steepest descent. In that case, note that the trajectory is also a geodesic \cite{Achucarro:2018vey, Calderon-Infante:2022nxb}. Away from the slow-roll approximation, or more generally at large speeds, we can imagine a more random trajectory, and it is unclear if the friction would be enough to regulate it. Interestingly, in \cite{Shiu:2023nph, Shiu:2023fhb}, it was shown that the asymptotic (fixed point) solution in a multi-exponential, canonical multifield, positive potential is not in slow-roll, but it still follows an ``accidental'' gradient flow. We see that under some conditions, the physical trajectory follows, at least asymptotically, a gradient flow, but it is not necessarily true in full generality (any potential, field metric, initial conditions).

We now consider that the trajectory follows a gradient flow. In addition, assume, as e.g.~in examples of Section \ref{sec:examples}, that the trajectory is along a field $\lambda(t)$, at least locally. As argued in \cite[Sec. 2.2.2]{Andriot:2023isc}, in that case, transverse directions to the trajectory are flat, i.e.~have a vanishing derivative of the potential. Indeed, consider a field $\eta$: one has
\beq
\del_{\eta} V = \frac{\del \varphi^i}{\del \eta} \del_{i} V = \frac{\del \varphi^i}{\del \eta} g_{ij} g^{jk} \del_k V = \frac{\del \overrightarrow{\varphi}}{\del \eta} \cdot \overrightarrow{\nabla} V \propto \frac{\del \overrightarrow{\varphi}}{\del \eta} \cdot \frac{\del \overrightarrow{\varphi}}{\del t} = \frac{\del \varphi^i}{\del \eta} g_{ij} \frac{\del \varphi^j}{\del \lambda}\, \dot{\lambda} = g_{\eta\, \lambda} \, \dot{\lambda} \ .
\eeq
If $\eta$ is transverse to the trajectory that is along $\lambda$, one has $g_{\eta\, \lambda} = 0$, from which we conclude $\del_{\eta} V = 0$.

The above shows that transverse fields, in a gradient flow situation, have vanishing derivatives and therefore that, at least in a canonical basis,
\beq
\nabla V = | \del_{\hat{\lambda}} V | \ .
\eeq
So a single field captures the full gradient. For example, the SdSC can then be phrased in terms of a single field slope ratio
\beq
\frac{| \del_{\hat{\lambda}} V |}{V} \geq \sqrt{2} \ .\label{SdSCsingle}
\eeq
We may then hope for an {\sl on-shell} characterisation of the potential via an asymptotic single field slope ratio $\frac{| \del_{\hat{\lambda}} V |}{V}$.\\

Let us summarize. Considering physical trajectories dictated by equations \eqref{Friedmann} - \eqref{fieldeq}, we have asked ourselves whether an {\sl on-shell} characterisation of the potential could be given in the asymptotics, especially one in terms of a single field slope ratio $\frac{| \del_{\hat{\lambda}} V |}{V}$. We have shown that under some conditions (e.g.~slow-roll or steepest descent), the asymptotic physical trajectory could be given by a gradient flow. In that case, directions transverse to the trajectory are flat, and the full gradient boils down, at least in a canonical basis, to a single field derivative. This situation offers a possibility of an asymptotic {\sl on-shell} characterisation in terms of a single field slope ratio: we already know one, given by the SdSC, as in \eqref{SdSCsingle}, and may wonder about others.

Interestingly, if the trajectory is following the steepest descent, there is less risk of hitting a vanishing bound $c=0$: indeed, if the single field slope ratio were to vanish along the steepest descent, this would imply that the potential is fully flat. One could then investigate on characterising steepest descents of stringy scalar potential: the universal aspect of this question is appealing. In \cite{Andriot:2023isc}, an example of such a trajectory was studied in depth, to conclude on a bump in the slope ratio $\frac{\nabla V}{V}$, analogously to the bump observed for the ratio with the species scale \cite{vandeHeisteeg:2023ubh}. In addition, the asymptotic value of the slope ratio was found in this example to be larger than $\sqrt{2}$, therefore leaving the SdSC as a better characterisation. It would be interesting to study other examples of steepest descent trajectories. As argued in the Introduction, including the K\"ahler moduli in the examples of \cite{Calderon-Infante:2022nxb} would also give a slope ratio larger than $\sqrt{2}$. More generally, as discussed above, since for a gradient flow the transverse fields are flat, this means evolving along a ridge,\footnote{While \cite[v1]{Cremonini:2023suw} suggested a 4d scalar potential with ridges, it turns out not to be the case. Note also the recent mention of ridges in the string-inspired 4d theory of \cite{Kallosh:2024pat}.} or along a valley as in an effectively single field models.\footnote{Having a true valley-like potential from string theory is difficult, because of the generically non-canonical fields, as the axions. If a set of fields is canonically normalised, then one can look at their Hessian to study the stability of the potential among them. Having a valley would then single out one direction out of all eigenvectors of the Hessian, while the others are stabilised. Since the fields are canonically normalised, one can rotate them to align them with the Hessian eigenvectors (similarly to neutrino flavor mixing, in order to align with mass eigenstates). However, when some of the fields are non-canonically normalised, as typically the axions from string theory, it is unlikely that the fields in the kinetic terms can be aligned with the Hessian eigenvectors, especially all along a valley and not just at a critical point. While the field space metric might be block diagonal between e.g.~saxions and axions, it is not necessarily the case of the Hessian, especially away from a critical point, or e.g.~if the potential is linear in the axions. It would still be interesting to look for more (true) valley examples. In that respect, see e.g.~the recent work \cite{Lanza:2024uis}.} It is not so easy to construct such examples from string theory, but it would be interesting to do so and evaluate the slope ratio at asymptotics, comparing it to the SdSC one.\\

Last but not least, one might be interested in getting accelerated expansion in a cosmological solution. This is realised if $-\dot{H}/H^2 \equiv \epsilon < 1$. As is well-known, for a slow-roll single field situation, or in the case of \cite{Shiu:2023nph, Shiu:2023fhb, Shiu:2024sbe} for the asymptotic solution,\footnote{The situation there is multifield (canonical), $V$ is a sum of positive exponentials, and $\epsilon_V \leq d-1$ to avoid a change of fixed point.} one gets $\epsilon \approx \epsilon_V \equiv (d-2)/4 \ (\nabla V /V )^2$. In that case, the question of acceleration gets related to the value of slope ratio, and famously, the SdSC may then forbid asymptotic acceleration (see however \cite{Andriot:2023wvg} for the case with spatial curvature). But more generally, $\epsilon$ and $\epsilon_V$ are different, and acceleration is governed by the former (see e.g.~\cite{Achucarro:2018vey}). This indicates that characterising the potential via its slope ratio may eventually be less relevant (see however \cite{Freigang:2023ogu}) to the question of accelerated expansion.

\vspace{0.4in}

\subsection*{Acknowledgements}

We thank N.~Cribiori, L.~Gallot, F.~Tonioni, D.~Tsimpis, T.~Weigand and T.~Wrase for helpful exchanges during the completion of this work. M.R.~and G.T.~are supported in part by the NSF grant PHY-2210271. M.R.~acknowledges the support of the Dr.~Hyo Sang Lee Graduate Fellowship from the College of Arts and Sciences at Lehigh University.

\newpage

\begin{appendix}

\section{Extra material for the string examples}\label{ap:extra}

\subsection{F-theory 4d scalar potential}\label{ap:CY}

Focusing on F-theory compactifications on CY${}_4$, we presented in Section \ref{sec:CY4}, following \cite{Calderon-Infante:2022nxb}, a general 4d scalar potential given by
\begin{align}
V =& \frac{1}{\text{vol}} \left( \right.A_{1} e^{- \sqrt{6} \hat{U}_{I} - \sqrt{2}\hat{\tau}_{I}}+A_{2} e^{- \sqrt{\frac{2}{3}} \hat{U}_{I} - \sqrt{2}\hat{\tau}_{I}}+A_{3} e^{ \sqrt{\frac{2}{3}} \hat{U}_{I} - \sqrt{2}\hat{\tau}_{I}}+A_{4} e^{\sqrt{6} \hat{U}_{I} - \sqrt{2}\hat{\tau}_{I}} \\
&
+A_{5} e^{-\sqrt{6} \hat{U}_{I} + \sqrt{2}\hat{\tau}_{I}}
+ A_{6} e^{- \sqrt{\frac{2}{3}} \hat{U}_{I} + \sqrt{2}\hat{\tau}_{I}}
+A_{7} e^{ \sqrt{\frac{2}{3}} \hat{U}_{I}
+ \sqrt{2}\hat{\tau}_{I}}+A_{8} e^{ \sqrt{6} \hat{U}_{I}
+ \sqrt{2}\hat{\tau}_{I}} +A_{9}-A_{\text{loc}}\left. \right)\,, \nonumber
\end{align}
where the axions polynomials $A_i$ are given as follows in terms of flux numbers
\be \label{coeff}
\begin{split}
A_{1} &= \left(f_{0}-h_{0} \tau_{R} \pm f_{1} U_{R} + h_{1} \tau_{R} U_{R} + \frac{1}{2} f^{1} U_{R}^{2}-\frac{1}{2} h^{1} \tau_{R} U_{R}^{2}\pm \frac{1}{6} f^{0} U_{R}^{3} + \frac{1}{6} h^{0} \tau_{R} U_{R}^{3}\right)^2  \\[2mm]
    A_{2} &= \left(f_{1} \pm h_{1} \tau_{R} \pm f^{1} U_{R} \mp h^{1} \tau_{R} U_{R} + \frac{1}{2} f^{0} U_{R}^{2} \pm \frac{1}{2} h^{0} \tau_{R} U_{R}^{2} \right)^{2} \\[2mm]
    A_{3} &= \left(f^{1} -h^{1} \tau_{R} \pm f^{0} U_{R}+h^{0} \tau_{R} U_{R}\right)^{2}  \\[2mm]
    A_{4} &= \left(f^{0} \pm h^{0} \tau_{R}\right)^{2}  \\[2mm]
    A_{5} &= \left(h_{0} - h_{1} U_{R} + \frac{1}{2}h^{1} U_{R}^{2} - \frac{1}{6}h^{0} U_{R}^{3}\right)^{2} \\[2mm]
    A_{6} &= \left(h_{1} - h^{1} U_{R} + \frac{1}{2} h^{0} U_{R}^{2}\right)^{2} \\[2mm]
    A_{7} &= \left(h^{1} - h^{0} U_{R}\right)^{2} \\[2mm]
    A_{8} &= (h^{0})^{2} \\[2mm]
    A_{9} &\in \mathbb{R}
   \end{split}
\ee
This scalar potential can be found in \cite[(3.16)]{Calderon-Infante:2022nxb} with the notations (there=here) $b+\mathrm{i} \, u = U, c + \mathrm{i} \, s = \tau$ and the following redefinitions $f_{6} = f_{0}, f_{4} = f_{1}, f_{2} = f^{1}, f_{0} = f^{0}, h_{0} = h_{0}, h_{1}=h_{1}, h_{2} = h^{1}, h_{3} = h^{0}$.

\subsection{Landau-Ginzburg model: a brief review}\label{ap:LG}

We provide here a brief review of Landau-Ginzburg models, in particular the one considered in Section \ref{sec:LG}. This string model will give rise to a 4d effective theory with a scalar potential, similar to the one obtained after a Calabi-Yau compactification, however without any K\"ahler modulus $(h^{(1,1)}=0)$. Since K\"ahler moduli capture volumes of internal cycles of K\"ahler manifolds, and there is none here, the Landau-Ginzburg model considered is part of what is called ``non-geometric compactifications'' of type IIB string theory \cite{Becker:2006ks, Becker:2007dn}.

Consider first the following two dimensional $\mathcal{N}=(2,2)$ supersymmetric field theory
\be \label{wsCFT}
S = \int \d^2 z \,\d^4 \theta\, \mathcal{K} \left(\{ x_i,\bar{x}_i\}\right) + \left(\int \d^2 z \,\d^2 \theta\, \mathcal{W}
\left(\{x_i\}\right) + c.c \right) \,,
\ee
where the $x_i$'s are chiral superfields and the worldsheet superpotential $\mathcal{W}$ is a quasi-homogeneous function of these chiral superfields. The theory is completely determined by the superpotential and under RG flow it is conjectured to flow to an IR fixed point \cite{Lerche:1989uy}. For a simple superpotential, $\mathcal{W}=x^{k+2}$, the central charge of the theory at the IR fixed point, given by $c=\frac{3 k}{k+2}$, matches that of a level-$k$ minimal model. For a compactification to four spacetime dimensions, the internal space CFT must have a central charge of $9$. Tensoring together multiple minimal models to form Gepner models then allows for CFTs with the appropriate central charge for a string background \cite{Gepner:1987qi}: indeed, we can consider here 9 copies with $k=1$. We then restrict to the $1^{9}$ Landau-Ginzburg model which has a Landau-Ginzburg description that is given by the following worldsheet superpotential
\be \label{LGSuperpotential}
\mathcal{W}=\sum_{i=1}^{9} x_{i}^{3}\,.
\ee
The above superpotential admits a discrete symmetry that is generated by the following action
\be \label{groupaction}
g:x_{i} \mapsto \omega\, x_{i}\, ,
\ee
where $\omega=e^{\frac{2 \pi \mathrm{i}}{3}}$. The corresponding $\mathbb{Z}_{3}$ orbifold ensures that the $U(1)_{R}$ charges are integral and that the spacetime theory has $\mathcal{N}=2$ supersymmetry. The Hodge numbers of the above Landau-Ginzburg model are given by
\be \label{DiamondNoOplane}
\begin{array}{ccccccc}
  &   &   & 1 &   &   &   \\
  &   & 0 &   & 0 &   &   \\
  & 0 &   & 0 &   & 0 &   \\
0 &   & 84 &   & 84 &   & 0 \\
  & 0 &   & 0 &   & 0 &   \\
  &   & 0 &   & 0 &   &   \\
  &   &   & 1 &   &   &   \\
\end{array}
\ee
We perform an orientifold that breaks half of the supersymmetry such that we arrive at an $\mathcal{N}=1$ supersymmetric theory in four dimensions. The orientifold is generated by the following involution
\be \label{orientifold}
\sigma:\left(x_{1},
x_{2}, x_{3}, x_{4}, x_{5}, x_{6}, x_{7}, x_{8}, x_{9}\right)
\mapsto
-\left(x_{2}, x_{1},x_{3}, x_{4}, x_{5}, x_{6}, x_{7}, x_{8}, x_{9}\right)\, ,
\ee
coupled to a worldsheet parity operation. It projects out some of the complex structure deformations and the corresponding Hodge diamond of the Landau-Ginzburg model is modified as shown below
\be
\begin{array}{ccccccc}
  &   &   & 1 &   &   &   \\
  &   & 0 &   & 0 &   &   \\
  & 0 &   & 0 &   & 0 &   \\
0 &   & 63 &   & 63 &   & 0 \\
  & 0 &   & 0 &   & 0 &   \\
  &   & 0 &   & 0 &   &   \\
  &   &   & 1 &   &   &   \\
\end{array}
\ee
We see that there is no K\"ahler modulus and that there are 63 complex structure moduli, 60 of them coming from the twisted sector; we will rather focus on the 3 untwisted complex structure moduli in the following. The above model is mirror dual to type IIA string theory compactified on $T^6/\mathbb{Z}_{3} \times \mathbb{Z}_{3}$. At the Fermat point in the complex structure moduli space, fluxes and orientifolds can be studied in the Landau-Ginzburg language \cite{Becker:2006ks, Becker:2022hse, Becker:2024ijy, Rajaguru:2024emw, Becker:2024ayh}. In addition to this, the large complex structure limit of the untwisted sector moduli has also been of interest \cite{Becker:2007dn, Ishiguro:2021csu, Bardzell:2022jfh, Cremonini:2023suw}. The marginal deformations of the worldsheet superpotential are identified with the complex structure moduli as shown below, where we only highlight the untwisted sector moduli
\be \label{Moduli}
\mathcal{W}=\sum_{i=1}^{9} x_{i}^{3} - \mathbf{U_{1}} x_{1} x_{2} x_{3} - \mathbf{U_{2}} x_{4} x_{5} x_{6}-\mathbf{U_{3}} x_{7} x_{8} x_{9}\,.
\ee

The resulting 4d $\mathcal{N}=1$ theory is determined by a K\"ahler potential and superpotential, $K, W$. In the large complex structure limit of the untwisted sector, $K,W$ depend only the three untwisted complex structure moduli, together with the axio-dilaton \cite{Becker:2007dn}. We give the resulting $K,W$ in Section \ref{sec:LG}: they are the starting point of our study on this 4d string effective theory. In particular, the resulting scalar potential can be derived \cite[(3.12)]{Cremonini:2023suw}, and is given as follows in terms of canonical fields
\ba \label{LGpot}
V &=& A_{1} \, e^{\sqrt{2}\,(-2 \hat{\tau}_{I}+\hat{U}_{1 I}+\hat{U}_{2 I}+\hat{U}_{3 I})} + A_{2} \, e^{\sqrt{2}\,(-2\hat{\tau}_{I}+\hat{U}_{1 I}+\hat{U}_{2 I}- \hat{U}_{3 I})}+A_{3} \, e^{\sqrt{2}\,(-2\hat{\tau}_{I}+\hat{U}_{1 I}-\hat{U}_{2 I}+ \hat{U}_{3 I})} \nonumber \\[2mm]
&&\nonumber + A_{4} \,  e^{\sqrt{2}\,(-2\hat{\tau}_{I}-\hat{U}_{1 I} + \hat{U}_{2 I} + \hat{U}_{3 I})} + A_{5} \,  e^{\sqrt{2}\,(-2\hat{\tau}_{I}+\hat{U}_{1 I} - \hat{U}_{2 I} - \hat{U}_{3 I})} + A_{6} \, e^{\sqrt{2}\,(-2\hat{\tau}_{I}-\hat{U}_{1 I} + \hat{U}_{2 I} - \hat{U}_{3 I})} \\[2mm]
&&\nonumber + A_{7} \, e^{\sqrt{2}\,(-2\hat{\tau}_{I}-\hat{U}_{1 I} - \hat{U}_{2 I} + \hat{U}_{3 I})} + A_{8} \, e^{\sqrt{2}\,(-2\hat{\tau}_{I}-\hat{U}_{1 I} - \hat{U}_{2 I} - \hat{U}_{3 I})} + A_{9} \,  e^{\sqrt{2}\,(-\hat{\tau}_{I}+\hat{U}_{1 I} + \hat{U}_{2 I} + \hat{U}_{3 I})} \\[2mm]
&&\nonumber + A_{10} \, e^{\sqrt{2}\,(-\hat{\tau}_{I}+\hat{U}_{1 I} + \hat{U}_{2 I} - \hat{U}_{3 I})} +A_{11} \, e^{\sqrt{2}\,(-\hat{\tau}_{I}+\hat{U}_{1 I} - \hat{U}_{2 I} + \hat{U}_{3 I})} + A_{12} \, e^{\sqrt{2}\,(-\hat{\tau}_{I}-\hat{U}_{1 I} + \hat{U}_{2 I} + \hat{U}_{3 I})} \\[2mm]
&&\nonumber + A_{13} \, e^{\sqrt{2}\,(-\hat{\tau}_{I}+\hat{U}_{1 I} - \hat{U}_{2 I} - \hat{U}_{3 I})} + A_{14} \, e^{\sqrt{2}(-\hat{\tau}_{I}-\hat{U}_{1 I} + \hat{U}_{2 I} - \hat{U}_{3 I})} + A_{15} \, e^{\sqrt{2}\,(-\hat{\tau}_{I}-\hat{U}_{1 I} - \hat{U}_{2 I} + \hat{U}_{3 I})} \\[2mm]
&&\nonumber + A_{16} \, e^{\sqrt{2}\,(-\hat{\tau}_{I}-\hat{U}_{1 I} - \hat{U}_{2 I} - \hat{U}_{3 I})} + A_{17} \, e^{\sqrt{2}(-\hat{\tau}_{I}+\hat{U}_{1 I})} + A_{18} \, e^{\sqrt{2}\,(-\hat{\tau}_{I}+\hat{U}_{2 I})} + A_{19} \, e^{\sqrt{2}\,(-\hat{\tau}_{I}+\hat{U}_{3 I})} \\[2mm]
&& + A_{20} \, e^{\sqrt{2}\,(-\hat{\tau}_{I}-\hat{U}_{1 I})} + A_{21} \, e^{\sqrt{2}(-\hat{\tau}_{I}-\hat{U}_{2 I})} +  A_{22} \, e^{\sqrt{2}(-\hat{\tau}_{I}-\hat{U}_{3 I})} + A_{23} \, e^{-\frac{3}{\sqrt{2}}\hat{\tau}_{I}}\,,
\ea
where the $A_{i}$ are polynomial functions of axions with flux numbers as coefficients.

Note that contrary to what was suggested in \cite[v1]{Cremonini:2023suw}, this 4d theory does not allow for field directions where the scalar potential forms a ridge. We will thus investigate further directions in the main text.

\subsection{Derivation of DGKT 4d theory}\label{ap:DGKT}

We derive here the 4d theory motivated and used in Section \ref{sec:DGKT}. We start with 10d massive type IIA supergravity in string frame, as described in \cite{DeWolfe:2005uu, Andriot:2022bnb, Andriot:2023fss}. The 10d spacetime is taken to be a direct product between a 4d spacetime and a 6d compact manifold, where the latter is the orbifold $T^6/\mathbb{Z}_3\times\mathbb{Z}_3$. The metric of the latter takes the following form
\begin{align}\label{metric}
\text{d}s^2=2(\kappa\sqrt{3})^{1/3}\sum_{i=1}^3\upsilon_i\left((\text{d}y^{2i-1})^2+(\text{d}y^{2i})^2\right)\,,
\end{align}
while the volume over $T^6/\mathbb{Z}_3\times\mathbb{Z}_3$ is given by $\text{vol}=\int \text{d}^6y\sqrt{g_6}=\kappa\upsilon_1\upsilon_2\upsilon_3$. The background fluxes are given by
\begin{align}
    F_4=\sqrt{2}\, e_i\, \tilde{w}^i\,,
    \quad\quad
    H_3=-p\, \beta_0\,,
    \quad\quad
    F_0=-\sqrt{2}\, m_0\,,
    \quad\quad
    F_2=0 \,,
\end{align}
where $\tilde{w}^i$ and $\beta_0$ are 4- and 3-forms related to relevant cycles invariant under the orbifold action; we refer to \cite{Andriot:2023fss} for more details on the above. Together with (smeared) $O_6$ orientifold planes, this is the 10d background on which one can perform the dimensional reduction to 4d. We do so following \cite{Andriot:2022bnb} (see also \cite{Andriot:2023fss}). After performing the following rescaling of the 4d metric, $g_{\mu\nu\,S} = (2\pi\sqrt{\alpha^{\prime}})^6 \frac{e^{2\phi}}{\text{vol}} g_{\mu\nu}$, we reach the 4d Einstein frame, and eventually obtain a 4d effective action of the form \eqref{action}, where the Planck mass is given by $M_p^2=(\pi\alpha^{\prime})^{-1}$.

The 4d scalar potential has the general form
\begin{align}\label{fullpotential}
    V
    =\frac{(2\pi\sqrt{\alpha^{\prime}})^6M_p^2}{2}\,\frac{e^{2\phi}}{\text{vol}}\Bigg(\frac{1}{2}\vert H_3\vert^2&-e^{\phi}\frac{T_{10}}{7}+\frac{e^{2\phi}}{2}\Bigg[\vert F_0\vert^2+\vert F_0B_2\vert^2+\left\vert F_4 +\frac{1}{2}F_0\,B_2\wedge B_2\right\vert^2 \nonumber \\
    &+\left\vert C_3\wedge H_3+F_4\wedge B_2 +\frac{1}{6}F_0\,B_2\wedge B_2\wedge B_2 \right\vert^2\Bigg] \Bigg) \,,
\end{align}
where we include here and below the quartic and sextic axionic terms, which is usually not done. $T_{10}$ corresponds to the $O_6$-plane contribution, related to the background fluxes by the tadpole cancelation condition or Bianchi identity, while $\phi$ is the dilaton. The 4d theory, and this scalar potential in particular, will depend on eight scalar fields (ignoring the twisted sector): the dilaton, the deformations of the internal space (sizes of the three $T^2$ tori), three axions $b_i$ coming from the Kalb-Ramond $B_2$ components and one axion $\xi$ from the $C_3$ field:
\begin{align}
    \phi\,,
    \quad
    \upsilon_{i=1,2,3}\,,
    \quad
    B_2=\sum_{i=1}^3b_i \, w^i\,,\quad
    C_3=\sqrt{2}\,\xi\,\alpha_0 \,,
\end{align}
where the 2- and 3-forms $w^i$ and $\alpha_0$ correspond again to relevant cycles (see  \cite{Andriot:2023fss}).

With the above definitions, one can compute the explicit form of the scalar potential. To that end, it is more convenient to introduce slightly different fields
\begin{align}\label{redef1}
    r_i^2=\sqrt{\frac{\vert m_0\vert}{E}}\, \vert e_i\vert\, \upsilon_i
    \,,\quad
    g=\frac{e^{\phi}}{\sqrt{\text{vol}}}\, \frac{1}{\vert p\vert}\sqrt{\frac{E}{\vert m_0\vert}}
    \,,\quad
    \tilde{b}_i= \sqrt{\frac{|m_0|}{2E}} \, \vert e_i\vert \, b_i
    \,,\quad
    \tilde{\xi}= \sqrt{\frac{|m_0|}{2E}} \, \vert p\vert \, \xi \,,
\end{align}
where $E=\vert e_1e_2e_3\vert/\kappa$. With respect to \cite{Andriot:2023fss}, we introduced an extra factor in the axions redefinition, which will allow to remove powers of $|m_0|/E$, otherwise appearing in their potential and kinetic terms. We are now ready to compute explicitly the scalar potential \eqref{fullpotential} in terms of the fields $r_i\,, g\,,\tilde{b}_i\,,\tilde{\xi}$: it is eventually given by
\begin{align}
 V\ \times &\ \frac{2}{(2\pi\sqrt{\alpha^{\prime}})^6M_p^2}\,\frac{ E^{3/2}}{p^4\vert m_0\vert^{5/2}} \label{potDGKTgen}\\
    =&\ \frac{1}{2} g^2\prod_{i=1}^3\frac{1}{r_i^2}
    -2\sqrt{2}g^3
    +g^4\sum_{i=1}^3r_i^4\prod_{j=1}^3\frac{1}{r_j^2}
    +g^4\prod_{i=1}^3r_i^2 \nn \\
    & + 2\frac{g^4}{\prod_{i=1}^3r_i^2}  \left[\prod_{i=1}^3r_i^4\sum_{i=1}^3\frac{\tilde{b}_i^2}{r_i^4}
    -2\prod_{i=1}^3r_i^2\sum_{i=1}^3s_ir_i^2 \frac{\tilde{b}_j}{r_j^2}\frac{\tilde{b}_k}{r_k^2} +\left(\tilde{\xi}+\sum_{i=1}^3s_i\tilde{b}_i\right)^2\right.\nn \\
    & \phantom{+2\frac{g^4}{\prod_{i=1}^3r_i^2}} \left. \ +
    2 \prod_{i=1}^3\tilde{b}_i^2\sum_{i=1}^3\frac{r_i^4}{\tilde{b}_i^2}
    -4 \left(\tilde{\xi}+\sum_{i=1}^3s_i\tilde{b}^i\right) \prod_{j=1}^3\tilde{b}_j
     + 4 \prod_{i=1}^3\tilde{b}_i^2 \right] \,,  \nn
\end{align}
where we used $pm_0 <0$ (see \cite{Andriot:2023fss}), and where $s_i$ refers to the following fluxes signs
\beq
s_i\equiv\text{sign}(m_0e_i)=\pm 1\,.
\eeq
The first line of the potential \eqref{potDGKTgen} corresponds to the saxionic part, while the next lines contain the quadratic, quartic and sextic axionic contributions, the last two being usually ignored.

Extremizing the potential, we recover the well-known negative minimum, with the following values
\begin{equation}\label{minima}
\begin{split}
   & \langle r_i\rangle^4=\frac{5}{3}\,,
    \quad\quad
    \langle g\rangle=\sqrt{\frac{27}{160}}\,,
    \quad\quad\langle \tilde{b}_i\rangle=0 \,,\quad \quad  \langle \tilde{\xi}\rangle= 0 \,,\\
    & \langle V\rangle=
    - \frac{(2\pi\sqrt{\alpha^{\prime}})^6 M_p^2}{2}\, \frac{p^4\vert m_0\vert^{5/2}}{E^{3/2}}\times \left(\frac{3}{5}\right)^{\frac{5}{2}}\frac{27}{2^8}  \ .
\end{split}
\end{equation}

Finally, for the kinetic terms, we simply follow \cite{Andriot:2023fss} together with the above definitions of the scalar fields, to get
\begin{align}\label{totkin}
    S_{kin}
    &=-\frac{M_p^2}{2}\int\text{d}^4x\sqrt{-g_{4}}\left(
    \frac{2}{g^2}(\partial g)^2
    +\sum_{i=1}^3\frac{2}{r_i^2}(\partial r_i)^2
    +\sum_{i=1}^3\frac{1}{r_i^4}(\partial  \tilde{b}_i)^2
    +2g^2(\partial \tilde{\xi})^2\right)\,.
\end{align}
The axionic kinetic terms are not, and cannot, be canonically normalized since any kind of redefinition will leave mixed terms. This will complicate the study of trajectories along them.

In Section \ref{sec:DGKT}, we restrict ourselves for simplicity to the case of isotropic fields, by taking
\beq
\forall i \ ,\quad r_i = r \ ,\quad \tilde{b}_i = \tilde{b} \ , \label{isotropy}
\eeq
which is realised at least at the vacuum. This can be understood as fixing the non-isotropic degrees of freedom to given values and only considering trajectories along isotropic directions. With the choice of isotropic fields \eqref{isotropy}, kinetic terms and potential simplify to \eqref{kiniso} and \eqref{potpot}.

\end{appendix}

\newpage

\addcontentsline{toc}{section}{References}

\providecommand{\href}[2]{#2}\begingroup\raggedright\endgroup

\end{document}